\documentclass[conference]{IEEEtran}

\usepackage{cite}
\usepackage{amsmath,amssymb,amsfonts}
\usepackage{graphicx}
\usepackage{textcomp}
\usepackage{xcolor}
\usepackage{listings}
\usepackage{booktabs}
\usepackage{tabularx}
\usepackage{threeparttable} % numbered table notes (\tnote / tablenotes)
\usepackage{placeins}       % \FloatBarrier -- keep floats within their section
\usepackage{float}          % [H] exact-here float placement
\usepackage{subcaption}
\newcolumntype{L}[1]{>{\raggedright\arraybackslash\hsize=#1\hsize}X}
\usepackage{hyperref}
\usepackage{bookmark}
\usepackage{cleveref}
\usepackage{tikz}
\usetikzlibrary{arrows.meta,positioning,calc,shapes.geometric,patterns}
\usepackage{tikzpeople} % `criminal' shape = the passive-adversary icon in
\usepackage{pgfplots}
\pgfplotsset{compat=1.18}
\usepackage{multirow}

\def\BibTeX{{\rm B\kern-.05em{\sc i\kern-.025em b}\kern-.08em
    T\kern-.1667em\lower.7ex\hbox{E}\kern-.125emX}}

\newcommand{\FreqCtrl}{393.0xx}
\newcommand{\FreqDDCH}{393.3xx}
\newcommand{\FreqTrafLo}{392.8xx}
\newcommand{\FreqTrafHi}{394.6xx}
\newcommand{\FreqRaster}{358.3xx}

\newcommand{\TetraNeighborRows}{%
    1  & 421\,1xx\,xxx & 84x & 6417 \\
    3  & 420\,9xx\,xxx & 83x & 6283 \\
    8  & 421\,6xx\,xxx & 86x & 6591 \\
    13 & 421\,6xx\,xxx & 86x & 6034 \\
    16 & 421\,2xx\,xxx & 85x & 6748 \\
    20 & 421\,3xx\,xxx & 85x & 6159 \\
    22 & 421\,0xx\,xxx & 84x & 6820 \\
    23 & 424\,5xx\,xxx & 98x & 6375 \\
    24 & 420\,9xx\,xxx & 83x & 6902 \\
    26 & 423\,0xx\,xxx & 92x & 6516 \\%
}

\ifdefined\WHreal \InputIfFileExists{frequencies.secret.tex}{}{}\fi

\begin{document}

\title{The Hidden Life of Public Safety Communications Signals: A Comparative Security Analysis of TETRA, TETRAPOL, and P25%\\[4pt]
}

\author{
  \IEEEauthorblockN{Larry Hernandez\quad Sergey Bratus}
  \IEEEauthorblockA{\textit{Dartmouth College}\\
    \{l.gr,\,Sergey.L.Bratus\}@dartmouth.edu}
}

\maketitle

% ------------------------------------------------------------------
\begin{abstract}

Public-safety agencies and critical infrastructure operators rely on trunked
land-mobile radio (LMR) systems,
based on P25, TETRA, and TETRAPOL. These systems are
expected to protect not just the content of a communication but the fact of it.
Yet LMR standards leave a stark gap between confidentiality of \emph{content} and 
of \emph{communication}: underneath an encrypted traffic plane, their signaling plane
is almost entirely in the clear. We probe the depth and impact of adversarial
inference from this exposed signaling.

Prior security analyses of these systems have concentrated on the content
plane---recovering encryption keys or capturing accidental cleartext. We show
that comparably sensitive information can be \emph{inferred from passively
observed signaling even if the content encryption were perfect}.

In particular, we show that across the trunked LMR standards, a passive, receive-only software-defined radio (SDR) observer can recover operationally sensitive network
topology and geography details, unit presence, mobility across cells and groups, organizational
structure, as well as operational security details such as special key domains and key-epoch rotation. This signaling-plane inference
reaches far
beyond the observer's direct area of reception, turning \emph{local} sniffing
into \emph{nationwide} network mapping capabilities that degrade or defeat LMR standards'
identity obfuscation through timing and association.
In the case of TETRAPOL, we demonstrate how inference and tracking of such signaling metadata and a standards-level confidentiality failure in emergency call handling enable unencrypted voice extraction.

Finally, we discuss potential countermeasures and mitigations, including specific recommendations for protecting inter-cell, base station and subscriber identities.

\end{abstract}

% ==================================================================
\section{Introduction}\label{sec:introduction}
% ------------------------------------------------------------------

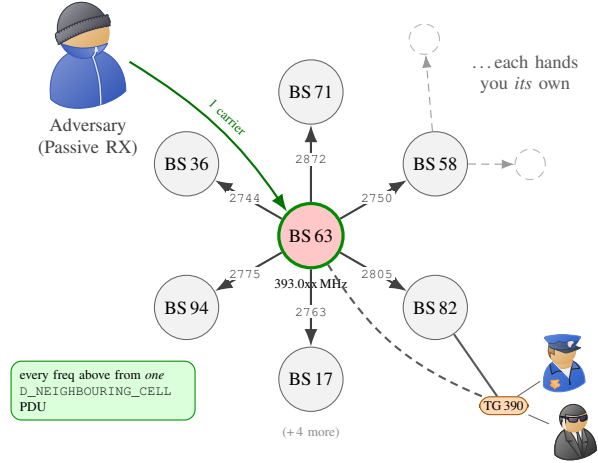
\begin{figure}[t]
  \centering
  \begin{tikzpicture}[
    font=\footnotesize, >=Latex,
    cell/.style={circle, draw=black!60, minimum size=0.85cm, inner sep=0pt, font=\scriptsize, align=center},
    serv/.style={cell, fill=red!22, draw=green!55!black, very thick},
    nbr/.style={cell, fill=gray!10, draw=black!60},
    ghost/.style={circle, draw=black!30, densely dashed, minimum size=0.4cm, inner sep=0pt},
    spoke/.style={->, thick, black!75},
    hop2/.style={->, densely dashed, black!40},
    chlbl/.style={font=\tiny\ttfamily, fill=white, inner sep=0.8pt, text=black!85},
    tg/.style={draw=orange!65!black, rounded corners, fill=orange!30, font=\tiny, inner sep=1.6pt},
    aff/.style={black!60, thin},
  ]
    \node[serv] (c) at (0,0) {BS\,63};
    \node[font=\tiny, below=0pt of c, align=center] {\FreqCtrl\,MHz};
    \node[nbr] (n1) at (90:1.9)  {BS\,71};
    \node[nbr] (n2) at (30:1.9)  {BS\,58};
    \node[nbr] (n3) at (-30:1.9) {BS\,82};
    \node[nbr] (n4) at (-90:1.9) {BS\,17};
    \node[nbr] (n5) at (210:1.9) {BS\,94};
    \node[nbr] (n6) at (150:1.9) {BS\,36};
    \draw[spoke] (c) -- (n1) node[chlbl,pos=0.55]{2872};
    \draw[spoke] (c) -- (n2) node[chlbl,pos=0.55]{2750};
    \draw[spoke] (c) -- (n3) node[chlbl,pos=0.55]{2805};
    \draw[spoke] (c) -- (n4) node[chlbl,pos=0.55]{2763};
    \draw[spoke] (c) -- (n5) node[chlbl,pos=0.55]{2775};
    \draw[spoke] (c) -- (n6) node[chlbl,pos=0.55]{2744};
    \node[ghost] (g1) at (60:3.0) {};
    \node[ghost] (g2) at (18:3.05) {};
    \draw[hop2] (n2) -- (g1);
    \draw[hop2] (n2) -- (g2);
    \node[font=\scriptsize, text=black!65, align=center] at (37:3.55) {\ldots each hands\\you \emph{its} own};
    \node[criminal, minimum size=1.0cm] at (-2.95,2.35) {};
    \node[font=\scriptsize, align=center, text=black!75] at (-2.95,1.3) {Adversary\\(Passive RX)};
    \draw[->, thick, green!45!black] (-2.4,2.3) to[bend left=12]
       node[font=\tiny, text=green!45!black, sloped, above, pos=0.5]{1 carrier} (c.north west);
    \node[draw=green!55!black, rounded corners, fill=green!14, font=\tiny, align=left,
          text width=2.2cm] at (-2.75,-2.05)
      {every freq above from \emph{one} \texttt{D\_NEIGHBOURING\_CELL} PDU};
    \node[font=\tiny, text=black!45] at (-90:2.6) {(+\,4 more)};
    \node[tg] (tg) at (2.55,-2.25) {TG\,390};
    \node[police, minimum size=0.55cm] (t1) at (3.35,-1.7) {};
    \node[maninblack, minimum size=0.55cm] (t2) at (3.5,-2.6) {};
    \draw[aff] (t1) -- (tg);
    \draw[aff] (t2) -- (tg);
    \draw[black!65, thick] (tg) -- (n3);
    \draw[black!65, thick, densely dashed] (tg) to[bend left=18] (c);
  \end{tikzpicture}
  \caption{\textbf{Self-guiding topology and affiliation inference.}
    One \texttt{D\_NEIGHBOURING\_CELL} broadcast from serving cell BS\,63 reveals each neighbor's identity and control-channel frequency. A passive observer can retune recursively to map the deployment. Talk group (TG) associations and cell transitions expose operational affiliation and mobility (\cref{sec:tp-neighbor}).}
  \label{fig:selfguiding}
\end{figure}

Public safety communications systems are a mission-critical component of modern law
enforcement and emergency response infrastructure worldwide. Their operators procure and rely
on them on the expectation of confidential operational communications: voice encryption is a
long-standing stated user need~\cite{NIJ2007VoiceEncryption,P25SteeringCommitteeCISA2023UserNeeds},
and national emergency and defense networks are fielded and promoted as
secure~\cite{Vogel2016ANTARES,CETSESIRDEE,Bundeswehr2021Tetrapol}. Although these systems use
different protocols in Europe (TETRAPOL, originally designed in France in the 1980s, and
being gradually succeeded by TETRA) and the US (Project 25, a.k.a. P25), they share
essential design similarities in modulation, error correction,
naming conventions of frames or PDUs, cryptographic schemes and algorithms, and so on.
As these standards and their deployments are long-lived,
often under decades-long servicing and maintenance contracts, weaknesses in their
designs create long-lived attack surfaces that must be understood and mitigated.

Seminal work by Glass et al., Clark et al., and Meijer et al.
\cite{glass2009sdr,glass2012insecurity,clark2011p25,meijer2023all} addressed aspects of
P25 and TETRA, noting their ad hoc designs that, e.g., eschewed cryptographic message
authentication and metadata encryption for the sake of availability in environments with interference or degraded RF conditions. Even so, the full scope of the cleartext signaling plane weaknesses of these
protocols---and especially of their hardware and software implementations---remains
under-explored.
In particular, the architectures of these standards invariably involve a split
between a \emph{signaling plane} (control channel) that carries metadata such as
registration, affiliation, and channel occupation (grants), and a \emph{traffic plane}
that carries the voice/data content. While the traffic plane may be encrypted, the
signaling plane remains exposed. We show that, through this unprotected, readily
accessible signaling plane, a passive, receive-only observer can infer a great deal of
sensitive information from the network, including its topology, operational security
properties (e.g., key rotation), churn and presence patterns, and even the organizational
structure of the subscriber agencies (\cref{sec:inference}).
We show that protocol-specific design choices allow a passive interception agent
to recursively bootstrap its understanding of a network from
a single cell's intercepted control carrier (Figure~\ref{fig:selfguiding}).

Moreover, support for vendor-defined applications, in combination with the complexity
of the underlying protocols, leads to practical weaknesses of the presumably protected
cryptographic layers, due to the relevant information spilling to the cleartext signaling
plane.
\begin{table*}[t]
  \centering
  \begin{threeparttable}
  \caption{Comparative matrix of public-safety digital radio standards}
  \label{tab:matrix}
  \small
  \begin{tabular}{@{}p{0.30\textwidth}p{0.13\textwidth}p{0.13\textwidth}p{0.13\textwidth}@{}}
    \toprule
    Exposure axis & TETRA & TETRAPOL & P25 \\
    \midrule
    Control channel encrypted?\tnote{3}      & no & no & no \\
    Identity/SSI enumeration        & partial\tnote{1} & yes\tnote{1} & yes \\
    Neighbor/topology broadcast    & full\tnote{2} & full\tnote{2} & yes \\
    Subscriber/TG census (passive)  & yes & yes & yes \\
    Cryptographic state/protocol visible?   & yes & yes & yes \\
    Idle-device beaconing leak      & yes~\cite{pfeifferAnalyzingTETRALocation2016} & yes & limited \\
    Content-crypto weakness         & TEA1~\cite{meijer2023all} & undisclosed\tnote{4} & DES/ADP~\cite{glass2012insecurity,clark2011p25} \\
    Weak-keystream content recovery & underexplored~\cite{meijer2023all} & undisclosed\tnote{4} & yes \\
    \bottomrule
  \end{tabular}
  \begin{tablenotes}[flushleft]
    \footnotesize
    \item[1] TETRA (SSI) and TETRAPOL (RTI/TTI) conceal identities with ephemeral IDs, but these persist within registration windows, so enumeration remains feasible.
    \item[2] Full neighbor graph: TETRAPOL via the \texttt{D\_NEIGHBOURING\_\allowbreak CELL} codop, TETRA via \texttt{D-NWRK-BROADCAST}.
    \item[3] \emph{No} holds for the broadcast/AACH blocks and the MAC header, which are never encrypted. TETRA C-plane signaling TM-SDUs \emph{may} be optionally encrypted under security class~2/3 (EN 300 392-2 \S21.4.3.1).
    \item[4] TETRAPOL air-interface encryption uses an undisclosed, proprietary algorithm; its strength and keystream recovery are out of scope.
  \end{tablenotes}
  \end{threeparttable}
\end{table*}

% ==================================================================
\section{Our Contributions}\label{sec:contributions}
% ------------------------------------------------------------------

In this paper we focus on the common weaknesses of the cleartext signaling planes
of LMR protocols. Our contributions are as follows.

\begin{enumerate}
  \item \textbf{A cross-protocol signaling-inference methodology.} We develop a
    metadata-first methodology that maps protocol-specific signaling from TETRA,
    TETRAPOL, and P25 into a common vocabulary of identity, topology, activity,
    mobility, affiliation, and cryptographic-state events
    (\cref{sec:methodology,sec:inference}).

  \item \textbf{WITCHHUNT (WH): an LMR whole-cell observatory.} We build WH, a
    real-time Rust framework for wideband carrier discovery, channelization,
    PHY/link decoding, transaction-state reconstruction, and normalized event
    extraction across multiple LMR protocols.
    (\cref{sec:method-pipeline}).

  \item \textbf{The first empirical TETRAPOL and TETRA cell security study, and a comparative
    deployment survey.} We passively surveyed operational TETRAPOL and TETRA
    networks, reconstructing neighbor
    graphs, frequency plans, subscriber and talk group (TG) populations, traffic
    patterns, and cryptographic key-index epochs
    (\cref{sec:tetrapol,sec:evaluation,sec:synthesis}).
  
    \begin{enumerate}
      \item \textbf{A standards-level confidentiality failure in TETRAPOL emergency handling.}
   The standard \emph{mandates} the Emergency Open Channel (ECH) to be unencrypted
   (PAS~0001-3-2 §4.4.33). Using a
   reverse-engineered RP-CELP vocoder, we recover a distress-button call's clear speech end to
   end, with no traffic-encryption key.
    \end{enumerate}

  \item \textbf{A platform for low-cost embedded experimentation.} We develop a
    TETRAPOL-capable embedded platform based on \texttt{MMDVM\_HS}---an open, low-cost
    ADF7021-based hotspot board---and use it to demonstrate decoding, topology
    discovery and applicability for near-field tracking through successful
    reception of downlink and uplink traffic.
\end{enumerate}

% ==================================================================
\section{Background: Trunked Public-Safety Digital Radio}\label{sec:background}
% ------------------------------------------------------------------
Trunked LMR systems pool RF channels among many users. A shared
control channel handles registration, TG affiliation, paging, and channel
grants. Assigned traffic channels then carry voice or data. This control/traffic
split improves spectrum use but creates the security seam studied here: payload
encryption may conceal content while the signaling needed to route it remains
observable. We discuss the origins and today's usage of these systems below.

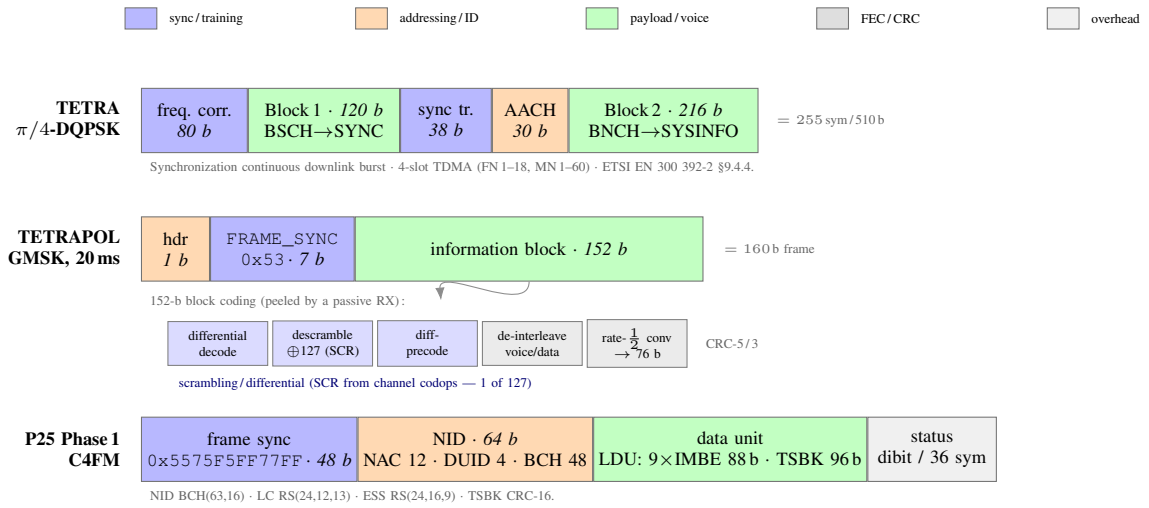
\begin{figure*}[t]
  \centering
  \begin{tikzpicture}[
    font=\footnotesize, >=Latex,
    fld/.style={draw=black!55, minimum height=0.85cm, inner xsep=2pt, font=\scriptsize, align=center},
    sync/.style={fld, fill=blue!28},
    addr/.style={fld, fill=orange!30},
    pay/.style={fld, fill=green!24},
    fec/.style={fld, fill=black!13},
    ovh/.style={fld, fill=black!6},
    rowlbl/.style={font=\scriptsize\bfseries, align=right, text width=2.7cm, anchor=east},
    sub/.style={font=\tiny, text=black!60, anchor=west},
    code/.style={draw=black!50, fill=black!8, minimum height=0.58cm, minimum width=1.32cm, inner xsep=1pt, font=\tiny, align=center},
    codehi/.style={code, fill=blue!14},
  ]
    \def\lx{2.75}
    \foreach \c/\t/\i in {sync/{sync\,/\,training}/0, addr/{addressing\,/\,ID}/1,
                          pay/{payload\,/\,voice}/2, fec/{FEC\,/\,CRC}/3, ovh/{overhead}/4}{
      \node[\c, minimum width=0.42cm, minimum height=0.28cm] (lg\i) at (\lx+\i*3.05,1.35) {};
      \node[right=1pt of lg\i, font=\tiny, anchor=west] {\t};
    }
    \node[rowlbl] at (\lx-0.15,0) {TETRA\\{\scriptsize$\pi/4$-DQPSK}};
    \node[sync, anchor=west, minimum width=1.4cm] (ta) at (\lx,0) {freq.\ corr.\\\textit{80 b}};
    \node[pay,  anchor=west, right=0pt of ta, minimum width=2.0cm] (tb) {Block\,1 \textit{· 120 b}\\{\scriptsize BSCH$\to$SYNC}};
    \node[sync, anchor=west, right=0pt of tb, minimum width=1.2cm] (ta2) {sync tr.\\\textit{38 b}};
    \node[addr, anchor=west, right=0pt of ta2, minimum width=1.0cm] (taach) {AACH\\\textit{30 b}};
    \node[pay,  anchor=west, right=0pt of taach, minimum width=2.5cm] (tc) {Block\,2 \textit{· 216 b}\\{\scriptsize BNCH$\to$SYSINFO}};
    \node[sub] at ($(tc.east)+(0.12,0)$) {$=255$\,sym\,/\,510\,b};
    \node[font=\tiny, text=black!60, anchor=west] at (\lx,-0.63)
       {Synchronization continuous downlink burst · 4-slot TDMA (FN\,1--18, MN\,1--60) · ETSI EN 300 392-2 \S9.4.4.};
    \node[rowlbl] at (\lx-0.15,-1.7) {TETRAPOL\\{\scriptsize GMSK, 20\,ms}};
    \node[addr, anchor=west, minimum width=0.9cm] (pa) at (\lx,-1.7) {hdr\\\textit{1 b}};
    \node[sync, anchor=west, right=0pt of pa, minimum width=1.9cm] (pb) {\texttt{FRAME\_SYNC}\\\texttt{0x53}\,\textit{· 7 b}};
    \node[pay,  anchor=west, right=0pt of pb, minimum width=4.6cm] (pc) {information block \textit{· 152 b}};
    \node[sub] at ($(pc.east)+(0.15,0)$) {$=160$\,b frame};
    \node[font=\tiny, text=black!60, anchor=west] (clbl) at (\lx,-2.4)
       {152-b block coding (peeled by a passive RX)\,:};
    \draw[->, black!45] (pc.south) to[out=-90,in=95] ($(clbl.east)+(0.35,0.02)$);
    \node[codehi, anchor=west] (s1) at (\lx+0.35,-2.95) {differential\\decode};
    \node[codehi, anchor=west, right=1.5pt of s1] (s2) {descramble\\$\oplus$127 (SCR)};
    \node[codehi, anchor=west, right=1.5pt of s2] (s3) {diff-\\precode};
    \node[code,   anchor=west, right=1.5pt of s3] (s4) {de-interleave\\voice/data};
    \node[code,   anchor=west, right=1.5pt of s4] (s5) {rate-$\tfrac12$ conv\\$\to$ 76 b};
    \node[sub] at ($(s5.east)+(0.12,0)$) {CRC-5\,/\,3};
    \node[font=\tiny, text=blue!45!black, anchor=north west] at ($(s1.south west)+(0,-0.04)$)
       {\,scrambling\,/\,differential (SCR from channel codops --- 1 of 127)};
    \node[rowlbl] at (\lx-0.15,-4.35) {P25 Phase\,1\\{\scriptsize C4FM}};
    \node[sync, anchor=west, minimum width=2.2cm] (qa) at (\lx,-4.35) {frame sync\\\texttt{0x5575F5FF77FF}\,\textit{· 48 b}};
    \node[addr, anchor=west, right=0pt of qa, minimum width=2.7cm] (qb)
       {NID \textit{· 64 b}\\{\scriptsize NAC 12 · DUID 4 · BCH 48}};
    \node[pay,  anchor=west, right=0pt of qb, minimum width=3.4cm] (qc)
       {data unit\\{\scriptsize LDU: 9$\times$IMBE 88\,b · TSBK 96\,b}};
    \node[ovh,  anchor=west, right=0pt of qc, minimum width=1.7cm] (qd) {status\\{\scriptsize dibit / 36 sym}};
    \node[font=\tiny, text=black!60, anchor=west] at (\lx,-4.97)
       {NID BCH(63,16) · LC RS(24,12,13) · ESS RS(24,16,9) · TSBK CRC-16.};
  \end{tikzpicture}
  \caption{\textbf{Shared on-air frame anatomy of TETRA, TETRAPOL, and P25.}
    The compared units combine synchronization (blue), addressing (orange), payload (green), and error protection (gray). The displayed TETRAPOL and P25 field sizes are decoder-verified, while the TETRA row shows the decoded \texttt{SYNC}/\texttt{BNCH} burst (\cref{sec:tetra-meta}). TETRAPOL's 152-bit block is expanded through its passive decoding chain.}
  \label{fig:frame-anatomy}
\end{figure*}

\subsection{TETRA}\label{sec:bg-tetra}
TETRA originated in an ETSI standardization effort begun in 1989 to provide
interoperable digital trunked radio for professional and public-safety users. ETSI
published the first standard in 1995, and TETRA subsequently developed into a
widely deployed voice-and-data system~\cite{etsi-tetra,esmaeilifar2025public}.
Its air interface uses four-slot time-division multiple access (TDMA) and supports both infrastructure-based
trunking and direct terminal-to-terminal operation. TETRA also provides
authentication, key management, and TEA-family air-interface encryption. We
discuss the relevant signaling exposure and TEA1 weakness in
\cref{sec:tetra-meta,sec:tetra-crypto}.

\subsection{TETRAPOL}\label{sec:bg-tetrapol}

\subsubsection{History of TETRAPOL}\label{sec:bg-tetrapol-history}

Developed by Matra in France in the 1980s, TETRAPOL is a proprietary frequency-division multiple access (FDMA)-based
digital radio protocol using Gaussian minimum-shift keying (GMSK) modulation. Its national-scale implementations
have primarily been used for national security and primary law-enforcement
purposes, including deployments in France, Spain, the Czech Republic, Slovakia,
Switzerland, and Germany~\cite{airbus-tetrapol,wikipedia-tetrapol}.

Large deployments also exist outside Europe. Mexico's implementation spans all
32 states and roughly 100{,}000 public-safety terminals~\cite{criticalcomms-mexico,rediris-mexico-history}.
Brazil deployed approximately 280 sites and tactical repeaters, including support
for security operations during the 2014 FIFA World Cup~\cite{airbus-integrapol,policeprofessional-worldcup}.
Together with smaller implementations elsewhere, the vendor-reported footprint is
roughly 80 networks across 34 countries~\cite{wikipedia-tetrapol}.

\subsubsection{Technical properties}\label{sec:bg-tetrapol-technical}

The normative reference for TETRAPOL is the set of Publicly Available Specification (PAS)
documents~\cite{TetrapolPubliclyAvailable2024}, primarily PAS 0001-1 (\emph{General Network Design}),
PAS 0001-3-1 (\emph{Air Interface Protocol; Air Interface Application Protocol}) and PAS 0001-3-3 (``Air Interface Protocol; Air Interface
Transport Protocol'').

The base station (\emph{Endpoint SwMI}) and subscribers (radios, \emph{RT applications}) communicate in a \emph{Base Network} (BN).
TSDUs are, in essence, the packets (or PDUs) passing over the air (\emph{PAS 0001-3-2 \S5.3,
93 Information Elements}~\cite{tetrapolPAS-0001-3-2}).

\subsubsection{Relationship to TETRA}

Despite the similarity in their names, TETRA and TETRAPOL even if often confused,
have little in common besides some of their design paradigms and nomenclature. The
design patterns do transfer between the standards, and so do their structural weaknesses,
as we demonstrate below. 

\subsection{P25}\label{sec:bg-p25}
APCO Project~25 (a.k.a. P25) is the TIA-102 suite for interoperable public-safety
land-mobile radio~\cite{tia102-p25}. Phase~1 uses continuous 4-level FM (C4FM) FDMA, while Phase~2 uses
two-slot TDMA. Both retain a trunked control plane that exposes network and
channel identifiers, TGs, unit addresses, grants, and affiliations.
Optional traffic encryption includes AES and legacy DES-OFB, alongside the
vendor-specific 40-bit ADP/RC4 mode~\cite{glass2012insecurity}, and deployments
can mix protected and clear calls~\cite{clark2011p25}. We analyze the resulting
metadata and fallback risks
in \cref{sec:p25-meta,sec:p25-crypto}.

DMR and related commercial LMR systems exhibit two relevant design overlaps,
namely exposed control signaling and optional per-call privacy. We mention
them only as corroborating context and do not analyze them separately, due to space constraints.

% ==================================================================
\section{Methodology: A Passive Signaling-Metadata Observatory}\label{sec:methodology}
% ------------------------------------------------------------------
\subsection{Capture and Demodulation Pipeline}\label{sec:method-pipeline}
WH ingests wideband IQ---from a file or a SoapySDR front end---and channelizes it
once through a shared polyphase filterbank into per-carrier baseband streams, so an
entire cell is observed in parallel rather than one channel at a time. Each control
carrier is decoded by an independent worker: GMSK demodulation and symbol-timing
recovery, then the TETRAPOL physical layer---frame synchronization, differential
decoding, descrambling, rate-$\tfrac12$ convolutional FEC, and de-interleaving---yielding
the \texttt{BCH}/\texttt{PCH}/\texttt{RCH}/\texttt{SDCH} logical channels. These feed
HDLC framing and per-connection link state, TSDU reassembly, and finally CODOP and
information-element (IE) dissection; a single serialized stage folds the results into
cell, neighbor, grant, and statistics state and emits records.

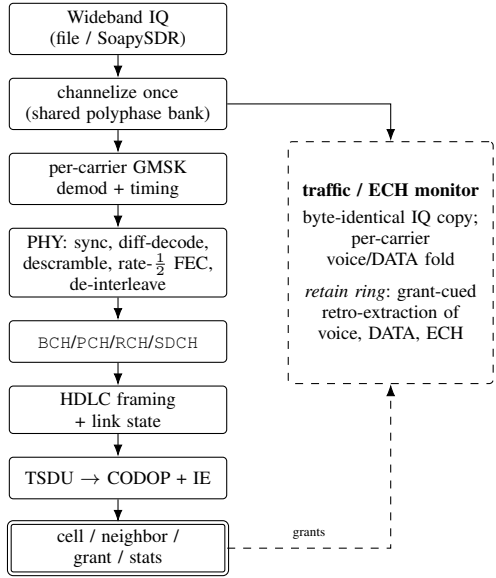
\begin{figure}[t]
  \centering
  \begin{tikzpicture}[
    font=\scriptsize,
    node distance=3mm and 8mm,
    box/.style={draw, rounded corners=1.5pt, align=center, inner sep=2.5pt,
                text width=27mm, minimum height=5.5mm},
    st/.style={box, double, double distance=1pt},
    mon/.style={draw, dashed, rounded corners=1.5pt, align=center, inner sep=3pt,
                text width=25mm},
    flow/.style={-{Latex[length=1.6mm]}},
  ]
    \node[box] (src)   {Wideband IQ\\(file / SoapySDR)};
    \node[box, below=of src]   (ch)    {channelize once\\(shared polyphase bank)};
    \node[box, below=of ch]    (demod) {per-carrier GMSK\\demod + timing};
    \node[box, below=of demod] (phy)   {PHY: sync, diff-decode,\\descramble, rate-$\tfrac12$ FEC,\\de-interleave};
    \node[box, below=of phy]   (lc)    {\texttt{BCH}/\texttt{PCH}/\texttt{RCH}/\texttt{SDCH}};
    \node[box, below=of lc]    (link)  {HDLC framing + link state};
    \node[box, below=of link]  (tsdu)  {TSDU $\to$ CODOP + IE};
    \node[st,  below=of tsdu]  (state) {cell / neighbor /\\grant / stats};
    \foreach \a/\b in {src/ch,ch/demod,demod/phy,phy/lc,lc/link,link/tsdu,tsdu/state}
      \draw[flow] (\a) -- (\b);
    \node[mon, right=8mm of phy, minimum height=32mm] (mon)
      {\textbf{traffic / ECH monitor}\\[2pt]
       byte-identical IQ copy;\\per-carrier voice/DATA fold\\[4pt]
       \emph{retain ring}: grant-cued\\retro-extraction of\\voice, DATA, ECH};
    \draw[flow] (ch.east) -| (mon.north);
    \draw[flow, dashed] (state.east) -| (mon.south)
      node[near start, above, font=\tiny] {grants};
  \end{tikzpicture}
  \caption{\textbf{WH capture-and-decode pipeline.} Wideband IQ is channelized once
    and every control carrier is decoded in parallel through the TETRAPOL PHY, logical
    channels, HDLC/link, and TSDU/CODOP/IE stages into serialized cell and grant state.
    A byte-identical copy feeds a traffic/ECH monitor whose retain ring, cued by
    recovered grants (dashed), retroactively extracts voice, DATA, and
    emergency-channel windows. The double border marks persistent state.}
  \label{fig:wh-pipeline}
\end{figure}

A parallel traffic monitor receives a byte-identical copy of the same IQ. Grants
recovered on the control plane cue it---and a separate retain ring---to retroactively
extract voice and DATA windows, including emergency-channel (ECH) traffic, from IQ
buffered \emph{before} the grant was observed, so WH captures a call it was never tuned
to in advance. Per-carrier decode runs on disjoint state in parallel, while grants,
topology, and retention are updated by serial, ordered owners, keeping the whole-cell
view consistent under load. The same channelize\,$\to$\,decode\,$\to$\,state architecture
is instantiated per protocol (for P25 via \texttt{dsd-neo}~\cite{dsdneo},
\cref{sec:evaluation}). \Cref{fig:wh-pipeline} summarizes the chain.

\subsection{Time-Domain Inference Primitives}\label{sec:method-primitives}

We build our protocol-agnostic analysis pipeline on the following
primitives that capture the trunked radio's signaling planes' observable properties,
events, and their temporal relationships.

\begin{enumerate}
  \item \textbf{Presence Detection}: The ability to detect the presence of a
  signal or device in the environment based on its unique characteristics (e.g.,
  metadata, such as subscriber and group IDs, or timing patterns).
  \item \textbf{Activity/Burst Clustering}: The ability to group similar
  signals or events based on their temporal characteristics and time-domain patterns.
  \item \textbf{Cross-Signal Correlation}: The ability to identify relationships
  between different signals or devices based on their properties (e.g., temporal and spectral)
  and their similarities.
  \item \textbf{Temporal and Co-occurrence Profiling}: The ability to analyze and model the
  temporal behavior of signals or devices through a defined \emph{observation window}.
\end{enumerate}

\subsection{Protocol-Specific Augmentation}\label{sec:method-augment}

We compose the generic primitives described in~\cref{sec:method-primitives} to capture
protocol-specific, semantically distinctive higher-level events, such as cryptographic key
rotation and over-the-air provisioning, subscriber or device registration, TG
affiliations, and emergency call response patterns.

These semantic events, in turn, are inputs rather than conclusions: the framework in
\cref{sec:inference} composes them across observation windows and cells into
security-relevant deductions. The protocol analyses in
\cref{sec:tetrapol,sec:tetra,sec:p25} identify the concrete signaling fields,
and \cref{sec:synthesis} compares their exposure across standards.

Protocol semantics also expose payload regularities, such as invariant
lengths, deterministic retransmissions, or predictable tunneled headers. Those
are content-analysis inputs rather than members of the signaling-inference
classes defined below.

\subsection{Decode Robustness as an Attack Enabler}\label{sec:method-robustness}

Robust reception expands the passive attack surface. Short uplink bursts,
including 30\,ms TETRAPOL segments,
interleaving, reassembly, and mixed voice and data frames complicate interception,
but the FEC and synchronization designed for weak subscriber links also benefit a
passive observer with similar capture geometry. The protocols further expose
coupled grant, channel, slot, call, and cryptographic states. Tracking these states
lets a resource-constrained observer follow active carriers and prioritize traffic
marked as clear, avoiding unnecessary processing of encrypted streams.

% ==================================================================
\section{Cross-Protocol Inference Framework}\label{sec:inference}
% ------------------------------------------------------------------

Building on the primitives and normalized events of
\cref{sec:method-primitives,sec:method-augment}, our framework joins observations
across time and cells. Although a single message may disclose only one identifier
or state, composing observations progressively exposes five classes of information. We define
those inference classes here, leaving protocol encodings to
\cref{sec:tetrapol,sec:tetra,sec:p25} and their comparison to
\cref{sec:synthesis}.

\subsection{Identity and Enumeration}\label{sec:inf-identity}

Network, site, subscriber, and TG identifiers support direct enumeration. Registration and authentication exchanges also bind stable identities to
temporary ones. Rotation therefore limits direct naming but does not necessarily
prevent linkage through persistent timing, cell, and group associations. The
protocol mechanisms are analyzed in
\cref{sec:tp-enum,sec:tetra-idcorr,sec:p25-meta}. Their cross-protocol
registration and affiliation vectors are compared in \cref{tab:inf-presence}.
We demonstrate the use of our inference pipeline in \cref{sec:evaluation}.

\subsection{Topology and Geography}\label{sec:inf-topology}
Serving- and neighbor-site broadcasts expose identifiers, channel plans, and
adjacent control frequencies. An observer can recursively retune to advertised
sites, turning one received control carrier into a self-guiding topology survey.
Reception from multiple locations further constrains coverage and geography. We
demonstrate this directly for TETRAPOL in \cref{sec:tp-neighbor}.
\cref{tab:inf-topology} enumerates the corresponding messages in each standard.

\subsection{Presence, Activity, and Mobility over Time}\label{sec:inf-presence}
Timestamped registration, affiliation, grant, and handover events establish
presence at a site. Repeated observation yields activity cadence and operational
tempo. Joining these events with the topology graph reveals movement between
cells. \Cref{tab:inf-presence} lists the relevant transactions, while
\cref{sec:eval-tracking} evaluates tracking across standards.

\subsection{Organizational-Structure Inference}\label{sec:inf-org}
Joining subscriber and TG identities with call endpoints, priorities,
paging, and emergency indicators produces a relationship graph of units and
groups without voice recovery. Inferring a group's real-world function requires
external labels, protocol-specific label leakage or contextual observation. The signaling nevertheless exposes
membership and interaction structure. \Cref{tab:inf-org} compares the carrying
messages, and \cref{sec:cs-p25} provides an empirical example.

\subsection{Operational Security and Confidentiality}\label{sec:inf-opsec}
Cleartext signaling of security class, cipher and key identifiers, or per-call
encryption state reveals where and when confidentiality is enabled. Longitudinal
observation distinguishes clear traffic, key-rotation epochs, and exceptional key
domains without recovering a key or payload. The concrete TETRAPOL and P25 cases
appear in \cref{sec:tp-ospec-patterns,sec:p25-opsec-algid}.
\Cref{tab:cross-protocol-synth-opsec} provides the cross-protocol mapping.

% ==================================================================
\section{TETRAPOL Security Analysis}\label{sec:tetrapol}
% ------------------------------------------------------------------

We surfaced several areas of interest and concerns, 
through development of our \texttt{WH} framework, analysis of both 
the public specifications (PAS) and the ways the deployments diverge from it, e.g., through
undocumented or reserved codops, and observation of two different deployments in
non-contiguous European countries.

\subsection{Control-Channel Signaling in the Clear}\label{sec:tp-clear}
Every TETRAPOL control channel broadcasts, in the clear, the static identity of
the network and the serving cell. From the signaling of a single measurement
site we recover the operator and network identifiers---a TETRAPOL
network, country~X, \texttt{system\_id}~61, network~5, at PAS specification
version~6---together with the serving-cell identity, base station BS~63 / RWS~1
(\texttt{BN}~88, \texttt{cell\_bn}~913). The same broadcasts expose the cell's
complete radio plan: a control channel at \FreqCtrl\,MHz, a dedicated data channel
(DDCH) at \FreqDDCH\,MHz (channel~2797), at least seven traffic channels spanning
\FreqTrafLo--\FreqTrafHi\,MHz, and the underlying channel raster (12.5\,kHz spacing on a
\FreqRaster\,MHz base). None of this requires decrypting a
single call: it is read in clear text from the control plane, and it
suffices to fingerprint the network and to tune a receiver to every carrier the
cell uses.

\subsection{Neighbor-Cell Broadcast: Whole-Network Topology Bootstrap}\label{sec:tp-neighbor}

TETRAPOL defines the \texttt{D\_NEIGHBOURING\_CELL} codop, as the cell's broadcast list of
adjacent cells (for mobile handover/cell-reselection). We decoded 10 neighbors---each
with a cell ID (BS/RSW) and the channel ID of that neighbor's control channel.

\begin{figure}[t]
  \centering
  \begin{tikzpicture}[
    font=\footnotesize, >=Latex,
    cell/.style={circle, draw=black!60, minimum size=0.66cm, inner sep=0pt, font=\tiny, align=center},
    serv/.style={cell, fill=gray!32, draw=green!55!black, very thick},
    nbr/.style={cell, fill=gray!10, draw=black!60},
    hex/.style={regular polygon, regular polygon sides=6, draw=black!50, minimum size=1.18cm, inner sep=0pt, font=\tiny},
    edge/.style={black!60, thick},
    explode/.style={->, very thick, green!45!black},
    panel/.style={font=\scriptsize\bfseries, text=black!75},
    tg/.style={draw=orange!65!black, rounded corners, fill=orange!30, font=\tiny, inner sep=2.8pt},
    aff/.style={black!60, thin},
  ]
    \node[criminal, minimum size=0.95cm] at (-3.05,2.3) {};
    \node[font=\scriptsize, align=center, text=black!95] at (-3.1,1.3) {Adversary\\(Passive RX)};
    \node[draw=green!55!black, rounded corners, fill=gray!5, align=left, font=\scriptsize,
          text width=3.0cm] (pdu) at (-0.55,1.9)
      {\textbf{\texttt{D\_NEIGHBOURING\_CELL}} {\scriptsize(one PDU/codop)}\\[2pt]
       \texttt{serving: BS\,63 @ \FreqCtrl}\\
       \texttt{BS\,71$\to$2872\ \ BS\,58$\to$2750}\\
       \texttt{BS\,82$\to$2805\ \ \ldots\,(10)}};
    \draw[->, thick, green!45!black] (-2.55,1.95) -- (pdu.west);
    \begin{scope}[shift={(-1.95,-1.15)}]
      \node[serv] (a47) at (0,0) {BS\,63};
      \node[nbr] (a20) at (90:1.05)  {71};
      \node[nbr] (a41) at (18:1.05)  {58};
      \node[nbr] (a48) at (-54:1.05) {82};
      \node[nbr] (a39) at (-126:1.05){17};
      \node[nbr] (a49) at (162:1.05) {94};
      \foreach \n in {a20,a41,a48,a39,a49} \draw[edge] (a47) -- (\n);
      \draw[edge] (a20) -- (a41);  \draw[edge] (a48) -- (a39);
      \node[tg] (tg) at (0,-1.65) {TG\,390};
      \node[police, minimum size=0.5cm] (t1) at (-0.9,-1.9) {};
      \node[maninblack, minimum size=0.5cm] (t2) at (0.9,-1.9) {};
      \draw[aff] (t1) -- (tg);  \draw[aff] (t2) -- (tg);
      \draw[edge] (tg) -- (a47);
      \draw[edge, densely dashed] (tg) to[bend right=22] (a48);
      \node[panel] at (-1.55,0.9) {Topology};
      \node[panel, text=black] at (0,-2.5) {+ affiliation};
    \end{scope}
    \begin{scope}[shift={(2.25,-1.15)}]
      \node[hex, fill=gray!38] (b47) at (0,0) {63};
      \node[hex, fill=gray!30] (b20) at (0,1.03) {71};
      \node[hex, fill=gray!20] (b41) at (0.9,0.51) {58};
      \node[hex, fill=gray!12] (b48) at (0.9,-0.51){82};
      \node[hex, fill=gray!20] (b39) at (0,-1.03) {17};
      \node[hex, fill=gray!12] (b49) at (-0.9,0.51){94};
      \node[panel] at (0,-1.75) {geography};
    \end{scope}
    \draw[explode] (pdu.south) to[out=-120,in=70]
       node[pos=0.6,left=12pt,font=\scriptsize,text=black!70]{cell IDs} (-1.95,0.1);
    \draw[explode] (pdu.south) to[out=-40,in=110]
       node[pos=0.55,right=1pt,font=\tiny,text=black!70]{} (2.25,0.05);
    \node[font=\tiny, text=black, align=center] at (2.25,-3.15)
      {each \texttt{ch} $=$ next cell's control channel};
  \end{tikzpicture}
  \caption{\textbf{One cleartext PDU, two inference paths.}
    A \texttt{D\_NEIGHBOURING\_CELL} broadcast exposes the neighbor graph and each adjacent cell's control-channel frequency, enabling recursive discovery. Combined with received signal strength, the same data supports coarse coverage inference. The topology panel also shows TG affiliation across cells. Cell identifiers are consistently pseudonymized across figures.}
  \label{fig:pdu-explode}
\end{figure}
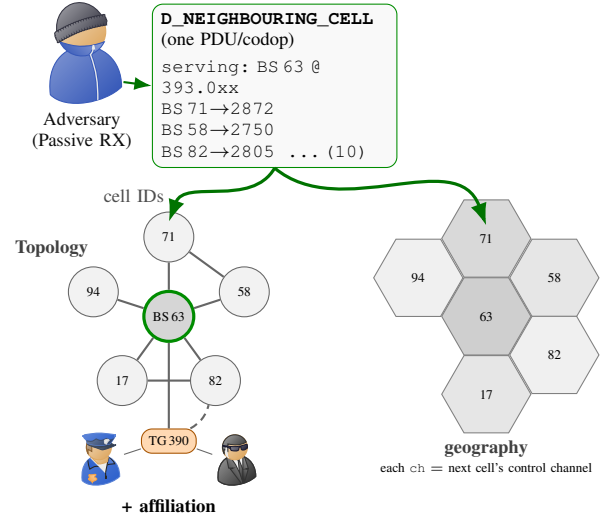

From a single carrier we bootstrap the structure of the whole deployed regional TETRAPOL
network: which cells exist, their IDs (e.g., BS 71/58/82/17/94/36/25/68/79/13), and how they're connected.

One PDU exposes the cell graph. This immediately enables:

\begin{enumerate}
   \item Frequency discovery $\rightarrow$ self-guiding survey. The neighbor channel ids (2872, 2750, 2805, 2763\ldots) are the RF control-channel frequencies of the neighbors.
  Instead of blind band-scanning, we
  decode one cell and it hands us the frequencies of all its neighbors---we walk the entire
  network cell-by-cell. This is the single most operationally valuable PDU for mapping an unknown
  deployment.

  \item Geolocation/coverage inference. Neighbor adjacency + measured signal strengths let us infer
  cell geography and coverage overlap---RF-planning-grade intelligence with no transmitter access.
  \item Mobility tracking. Combined with subscriber registrations, the neighbor graph lets us reason
  about which cells a unit can move between.
\end{enumerate}

We thus observe high-value metadata that leaks structural network intelligence with zero decryption, at a minimal
cost of decoding a single PDU.

Further composition and confirmation comes from secondary and tertiary functional codops, like
\texttt{D\_GROUP\_\allowbreak ACTIVATION} (e.g., a call exists in group 390 on channel 2878), explicitly tying signaling to topology.

\subsection{Subscriber, TG, and Registration Enumeration}\label{sec:tp-enum}

PAS defines registration as two sequential phases. An RT first draws a random
terminal identity (\texttt{RTI}). The SwMI replies to that address with the codop-less
\texttt{D\_TTI\_ASSIGNMENT}, then opens a level-2 connection using the assigned
\texttt{TTI}. The ensuing \texttt{U\_REGISTRATION\_REQ} carries the permanent
\texttt{HOST\_ADR} and \texttt{SERIAL\_NB}. \texttt{D\_REGISTRATION\_ACK} confirms
the \texttt{TTI} and repeats \texttt{HOST\_ADR}. Downlink observation can therefore
bind \texttt{RTI}$\leftrightarrow$\texttt{TTI}$\leftrightarrow$\texttt{HOST\_ADR},
while serial-number linkage requires the uplink~\cite{tetrapolPAS-0001-3-1,tetrapolPAS-0001-3-2}.

Clear group signaling supports a separate census. \texttt{D\_GROUP\_ACTIVATION}
and \texttt{D\_GROUP\_PAGING} expose group identity and activity, but do not by
themselves identify membership. \texttt{D\_REGISTRATION\_ACK} may also carry
\texttt{GROUP\_ID}, while the uplink \texttt{U\_ATTACH} reports the selected group
when a terminal powers on or changes selection. Those transactions directly bind a
terminal to a group when captured. Traffic reflected from the downlink towards subscriber terminals also contributes to the inference capability, in responses containing non-broadcast addressing. Therefore, a missed uplink codop may still result in reflected activity observable in the downlink in close temporal correlation.

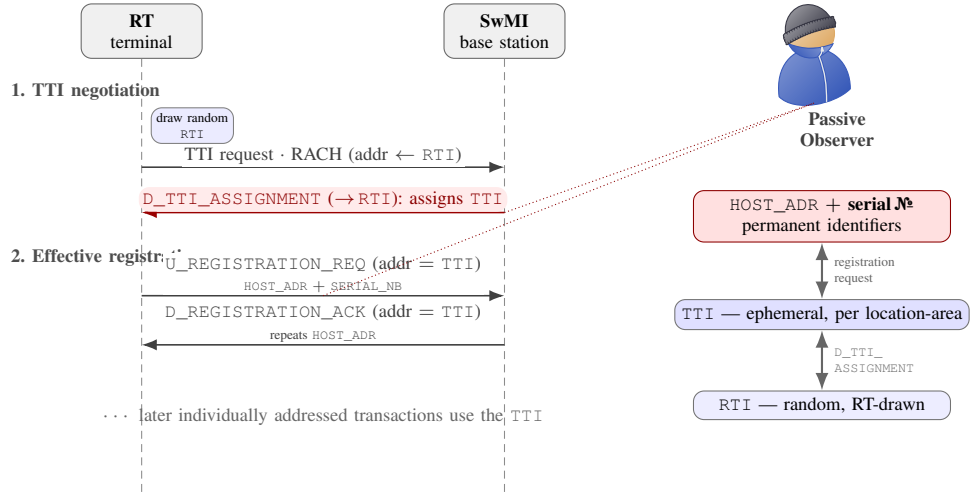
\begin{figure*}[t]
  \centering
  \begin{tikzpicture}[
    font=\footnotesize, >=Latex,
    actor/.style={draw=black!60, fill=black!6, rounded corners, align=center, font=\scriptsize\bfseries, minimum height=0.72cm, minimum width=1.6cm},
    life/.style={densely dashed, black!45},
    msg/.style={->, semithick, black!75},
    mlbl/.style={font=\scriptsize, fill=white, inner sep=1pt},
    leak/.style={->, semithick, red!60!black},
    leaklbl/.style={font=\scriptsize, fill=red!8, inner sep=1.5pt, text=red!55!black, rounded corners},
    phase/.style={font=\scriptsize\bfseries, text=black!70, anchor=west},
    selfbox/.style={draw=black!50, fill=blue!8, rounded corners, font=\tiny, inner sep=2pt, align=center},
    idbox/.style={draw=black!55, rounded corners, align=center, font=\scriptsize, inner sep=2.5pt, minimum width=3.4cm},
  ]
    \def\rt{2.0}\def\sw{6.8}
    \node[actor] (rt) at (\rt,3.35) {RT\\{\scriptsize\mdseries terminal}};
    \node[actor] (sw) at (\sw,3.35) {SwMI\\{\scriptsize\mdseries base station}};
    \draw[life] (rt) -- (\rt,-2.75);
    \draw[life] (sw) -- (\sw,-2.75);
    \node[phase] at (0.15,2.55) {1.\ TTI negotiation};
    \node[selfbox, anchor=west] at ($(\rt,2.1)+(0.12,0)$) {draw random\\\texttt{RTI}};
    \draw[msg] (\rt,1.55) -- node[mlbl,above]{TTI request · RACH (addr $\leftarrow$ \texttt{RTI})} (\sw,1.55);
    \draw[leak] (\sw,0.95) -- node[leaklbl,above]{\texttt{D\_TTI\_ASSIGNMENT} ($\to$\,\texttt{RTI}): assigns \texttt{TTI}} (\rt,0.95);
    \node[phase] at (0.15,0.35) {2.\ Effective registration};
    \draw[msg] (\rt,-0.15) -- node[mlbl,above,align=center]{\texttt{U\_REGISTRATION\_REQ} (addr $=$ \texttt{TTI})\\{\tiny \texttt{HOST\_ADR} $+$ \texttt{SERIAL\_NB}}} (\sw,-0.15);
    \draw[msg] (\sw,-0.8) -- node[mlbl,above,align=center]{\texttt{D\_REGISTRATION\_ACK} (addr $=$ \texttt{TTI})\\{\tiny repeats \texttt{HOST\_ADR}}} (\rt,-0.8);
    \node[font=\scriptsize, text=black!60] at ($(\rt,-1.75)!0.5!(\sw,-1.75)$) {$\cdots$ later individually addressed transactions use the \texttt{TTI}};
    \node[criminal, minimum size=0.95cm] at (10.9,3.0) {};
    \node[font=\scriptsize\bfseries, align=center, text=black!75] at (11.2,2.05) {Passive\\Observer};
    \draw[densely dotted, red!55!black] (\sw,0.95) -- (10.9,2.4);
    \draw[densely dotted, red!55!black] (4.4,-0.15) -- (10.9,2.4);
    \node[idbox, fill=red!12, draw=red!55!black] (serial) at (11.0,0.9) {\texttt{HOST\_ADR} $+$ \textbf{serial\,\textnumero}\\permanent identifiers};
    \node[idbox, fill=blue!12] (tti) at (11.0,-0.4) {\texttt{TTI} --- ephemeral, per location-area};
    \node[idbox, fill=blue!7] (rti) at (11.0,-1.6) {\texttt{RTI} --- random, RT-drawn};
    \draw[<->, thick, black!60] (serial) -- node[right=1pt,font=\tiny,align=left]{registration\\request} (tti);
    \draw[<->, thick, black!60] (tti) -- node[right=1pt,font=\tiny,align=left]{\texttt{D\_TTI\_}\\\texttt{ASSIGNMENT}} (rti);
  \end{tikzpicture}
  \caption{\textbf{TETRAPOL registration links ephemeral and permanent identities.}
    \texttt{D\_TTI\_ASSIGNMENT} binds the terminal-drawn \texttt{RTI} to an ephemeral \texttt{TTI}. The ensuing \texttt{U\_REGISTRATION\_REQ} carries \texttt{HOST\_ADR} and \texttt{SERIAL\_NB}, while \texttt{D\_REGISTRATION\_ACK} repeats \texttt{HOST\_ADR}. Downlink observation links \texttt{RTI}, \texttt{TTI}, and \texttt{HOST\_ADR}. Linking the serial number requires uplink capture. The analogous TETRA binding appears in \cref{fig:tetra-reg}.}
  \label{fig:tetrapol-reg}
\end{figure*}

\subsection{Connection-Oriented Data Sessions and SDS}\label{sec:tp-data}

PAS reserves connection-oriented transport on the SDCH for application exchanges,
with a new transport connection for each transaction. \texttt{D\_DATA\_REQUEST}
announces a downlink data transfer, \texttt{D\_CONNECT\_CCH} or
\texttt{D\_CONNECT\_DCH} selects its bearer, and \texttt{D\_DCH\_OPEN} opens
the data channel. Addressed link framing and clear setup fields disclose the endpoint,
priority, transport mode, key references, selected bearer, timing, and traffic volume
even when the application body remains opaque~\cite{tetrapolPAS-0001-3-1,tetrapolPAS-0001-3-2,tetrapolPAS-0001-3-3}.

PAS~0001-13-2 defines a
structured Submit/Delivery Protocol header containing, among other fields,
\texttt{CHIFF}, priority, application type, message type, encoding, and participant addresses. This reveals who exchanged data, when, how much, and which application family was used~\cite{tetrapolPAS-0001-13-2}.

\subsection{Observable Operational Security Patterns: Key Use and Rotation}\label{sec:tp-ospec-patterns}

PAS~0001-3-2~\cite{tetrapolPAS-0001-3-2}, §5.3.41, defines the \texttt{KEY\_REFERENCE} octet as a \texttt{KEY\_TYPE} (high nibble) and \texttt{KEY\_INDEX} (low nibble) pair, which
describe the cryptographic type and slot used for the encryption of the voice/data traffic. The
same octet rides the channel-activation and call-setup codops---\texttt{D\_GROUP\_ACTIVATION}
(§4.4.44), \texttt{D\_ECH\_ACTIVATION} (§4.4.33), \texttt{D\_OC\_ACTIVATION} (§4.4.60),
\texttt{D\_CALL\_START} (§4.4.14), and \texttt{D\_CALL\_CONNECT} (§4.4.12)---so the key in force is
\emph{announced on the control channel at every call setup.} These
fields are readily observable in the clear, allowing precise tracking of the key rotation
schedule. In two separate TETRAPOL cells in active deployment, we were able to reliably
discriminate between key epochs, based on the shift of the key index at specific times of day.

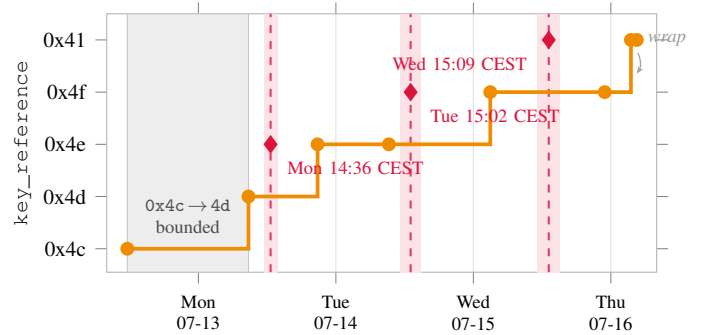
\begin{figure}[H]
\centering
\definecolor{keyorange}{RGB}{240,140,0}
\definecolor{keycrimson}{RGB}{220,20,60}
\begin{tikzpicture}
\begin{axis}[
    width=\columnwidth, height=5.0cm,
    xmin=8, xmax=104,
    ymin=-0.45, ymax=4.5,
    xtick={0,24,48,72,96,120,144},
    xticklabels={{Sun\\07-12},{Mon\\07-13},{Tue\\07-14},{Wed\\07-15},{Thu\\07-16},{Fri\\07-17},{Sat\\07-18}},
    xticklabel style={align=center, font=\scriptsize, yshift=-1pt},
    ytick={0,1,2,3,4},
    yticklabels={0x4c,0x4d,0x4e,0x4f,0x41},
    yticklabel style={font=\footnotesize},
    ylabel={\footnotesize \texttt{key\_reference}},
    ylabel style={yshift=-2pt},
    xmajorgrids=true,
    grid style={gray!25, line width=0.3pt},
    tick align=outside,
    axis line style={gray!60},
    clip mode=individual,
    enlargelimits=false,
]
\filldraw[fill=gray!14, draw=gray!45, line width=0.4pt]
    (axis cs:11.5889,-0.45) rectangle (axis cs:32.7436,4.5);
\node[gray!60!black, font=\scriptsize, align=center] at (axis cs:22.17,0.62)
    {$\mathtt{0x4c}\!\to\!\mathtt{4d}$\\bounded};
\fill[keycrimson!12] (axis cs:35.3958,-0.45) rectangle (axis cs:37.8083,4.5);
\fill[keycrimson!12] (axis cs:59.2775,-0.45) rectangle (axis cs:62.8178,4.5);
\fill[keycrimson!12] (axis cs:83.1644,-0.45) rectangle (axis cs:87.16,4.5);
\draw[keycrimson, dashed, line width=0.8pt, opacity=0.8] (axis cs:36.6022,-0.45) -- (axis cs:36.6022,4.5);
\draw[keycrimson, dashed, line width=0.8pt, opacity=0.8] (axis cs:61.0478,-0.45) -- (axis cs:61.0478,4.5);
\draw[keycrimson, dashed, line width=0.8pt, opacity=0.8] (axis cs:85.1622,-0.45) -- (axis cs:85.1622,4.5);
\addplot[const plot mark left, keyorange, line width=1.4pt,
         mark=*, mark size=1.9pt, mark options={fill=keyorange, draw=keyorange}]
    table[col sep=comma, x=x_hours, y=keyref_ord] {data/tp-keyday/keyday_series.csv};
\addplot[only marks, mark=diamond*, mark size=3.2pt,
         mark options={fill=keycrimson, draw=keycrimson}]
    table[col sep=comma, x=x_detect_hours, y=to_ord] {data/tp-keyday/keyday_rotations.csv};
\node[keycrimson, font=\scriptsize, anchor=west] at (axis cs:37.9,1.55) {Mon 14:36 CEST};
\node[keycrimson, font=\scriptsize, anchor=west] at (axis cs:62.9,2.55) {Tue 15:02 CEST};
\node[keycrimson, font=\scriptsize, anchor=east] at (axis cs:83.0,3.55) {Wed 15:09 CEST};
\node[gray!80, font=\scriptsize\itshape, anchor=west] at (axis cs:100.9,4.0) {wrap};
\draw[gray!70, -{Stealth[length=3pt]}, line width=0.5pt]
    (axis cs:100.7,3.75) to[bend left=30] (axis cs:100.5,3.35);
\end{axis}
\end{tikzpicture}
\caption{\textbf{Daily network-synchronous key-day roll} of an operational TETRAPOL cell
over four days.}
\label{fig:tp-keyday}
\end{figure}

In one observed deployment (\cref{fig:tp-keyday}), the index advances sequentially
($\mathtt{0x4c}\!\to\!\mathtt{4d}\!\to\!\mathtt{4e}\!\to\!\mathtt{4f}$, then wraps to
$\mathtt{0x41}$). Crimson diamonds mark each roll's first-detect estimate with its
$\pm$10\% bounding band and local time---the rolls cluster at $\approx$13:00\,UTC
($\approx$15:00\,CEST) daily. The grey span is the bounded
$\mathtt{0x4c}\!\to\!\mathtt{4d}$ window that closed before the capture began. 

More interesting is the fact that emergency calls (e.g., driven by the \texttt{D\_ECH\_} codops)
were persistently non-encrypted in one deployment (\cref{sec:evaluation}).

% ==================================================================
\section{TETRA Security Analysis}\label{sec:tetra}
% ------------------------------------------------------------------
\subsection{Signaling-Plane Metadata Leakage}\label{sec:tetra-meta}

Similarly to TETRAPOL (\cref{sec:tp-clear}), TETRA performs the core control and signaling plane
operations in the clear. Two distinct PDU types (\texttt{SYSINFO}, \texttt{D-NWRK-BROADCAST}) convey
network topology information.

A single SYNC+SYSINFO capture is sufficient to reliably determine the cell identity, encryption status,
spectrum topology (frequencies of operation) and other details, from a single demodulated and decoded
carrier.

A subsequent \texttt{D-NWRK-BROADCAST} reveals the neighboring cells and their downlink
carrier frequencies and respective IDs.

\begin{table}[h]
  \centering
  \caption{Neighbor cells advertised in \texttt{D-NWRK-BROADCAST} on the
    surveyed TETRA cell (10 of 25 observed). Downlink carriers and carrier numbers
    are masked to band level. Location Areas are consistent anonymized identifiers.}
  \label{tab:tetra-neighbors}
  \footnotesize
  \begin{tabular}{@{}rrrr@{}}
    \toprule
    \textbf{Cell id} & \textbf{DL carrier (Hz)} & \textbf{Carrier no.} & \textbf{LA} \\
    \midrule
    \TetraNeighborRows
    \bottomrule
  \end{tabular}
\end{table}

Table~\ref{tab:tetra-neighbors} shows the extracted frequency plan for a surveyed TETRA cell, from a
single cell. Each row hands a passive observer an adjacent cell's downlink carrier and Location Area,
bootstrapping the network graph from a single carrier.

\subsection{Intra-cell or External Traffic Reflection}\label{sec:tetra-reflection}

In deployments that utilize IP gateways or tunneling---e.g., TETRAPOL-TETRA interconnects---an
additional attack surface exists: any reflected traffic specific to one cell or system transmitted
through another---e.g., a cell deployed in a different location---may expose the source cell's location and affiliation to inference.

\subsection{Identity Correlation: Ephemeral Aliases, Event Labels, and the TEI}\label{sec:tetra-idcorr}

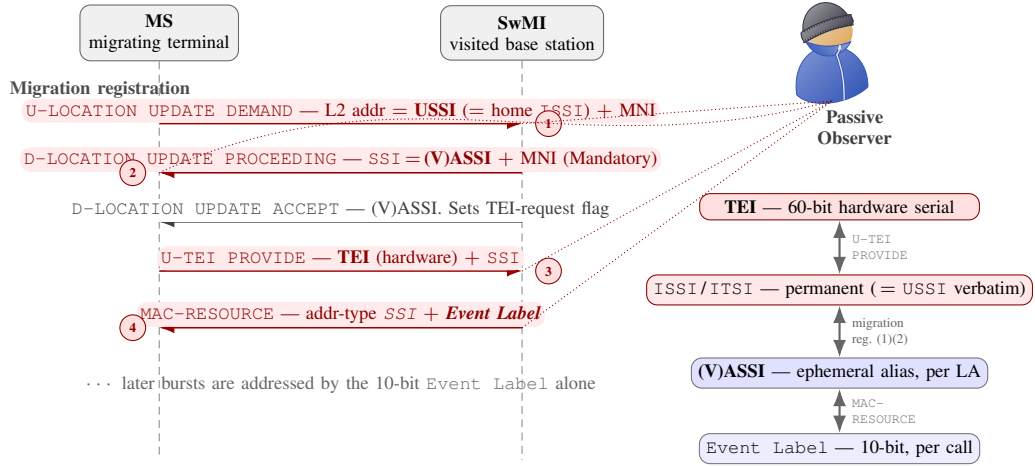
\begin{figure*}[t]
  \centering
  \begin{tikzpicture}[
    font=\footnotesize, >=Latex,
    actor/.style={draw=black!60, fill=black!6, rounded corners, align=center, font=\scriptsize\bfseries, minimum height=0.72cm, minimum width=1.7cm},
    life/.style={densely dashed, black!45},
    msg/.style={->, semithick, black!75},
    mlbl/.style={font=\scriptsize, fill=white, inner sep=1pt},
    leak/.style={->, semithick, red!60!black},
    leaklbl/.style={font=\scriptsize, fill=red!8, inner sep=1.5pt, text=red!55!black, rounded corners},
    phase/.style={font=\scriptsize\bfseries, text=black!70, anchor=west},
    tag/.style={circle, draw=red!55!black, fill=red!12, inner sep=0.5pt, font=\tiny\bfseries, text=red!55!black, minimum size=0.34cm},
    idbox/.style={draw=black!55, rounded corners, align=center, font=\scriptsize, inner sep=2.5pt, minimum width=3.7cm},
  ]
    \def\ms{2.2}\def\sw{7.0}
    \node[actor] (ms) at (\ms,3.15) {MS\\{\scriptsize\mdseries migrating terminal}};
    \node[actor] (sw) at (\sw,3.15) {SwMI\\{\scriptsize\mdseries visited base station}};
    \draw[life] (ms) -- (\ms,-2.5);
    \draw[life] (sw) -- (\sw,-2.5);
    \node[phase] at (0.1,2.4) {Migration registration};
    \draw[leak] (\ms,1.95) -- node[leaklbl,above]{\texttt{U-LOCATION UPDATE DEMAND} --- L2 addr $=$ \textbf{USSI} ($=$ home \texttt{ISSI}) $+$ MNI} (\sw,1.95);
    \node[tag] at (\sw+0.35,1.95) {1};
    \draw[leak] (\sw,1.30) -- node[leaklbl,above]{\texttt{D-LOCATION UPDATE PROCEEDING} --- \texttt{SSI}\,$=$\,\textbf{(V)ASSI} $+$ MNI (Mandatory)} (\ms,1.30);
    \node[tag] at (\ms-0.35,1.30) {2};
    \draw[msg] (\sw,0.65) -- node[mlbl,above]{\texttt{D-LOCATION UPDATE ACCEPT} --- (V)ASSI. Sets TEI-request flag} (\ms,0.65);
    \draw[leak] (\ms,0.0) -- node[leaklbl,above]{\texttt{U-TEI PROVIDE} --- \textbf{TEI} (hardware) $+$ \texttt{SSI}} (\sw,0.0);
    \node[tag] at (\sw+0.35,0.0) {3};
    \draw[leak] (\sw,-0.75) -- node[leaklbl,above]{\texttt{MAC-RESOURCE} --- addr-type \emph{\texttt{SSI} $+$ \textbf{Event Label}}} (\ms,-0.75);
    \node[tag] at (\ms-0.35,-0.75) {4};
    \node[font=\scriptsize, text=black!60] at ($(\ms,-1.5)!0.5!(\sw,-1.5)$) {$\cdots$ later bursts are addressed by the 10-bit \texttt{Event Label} alone};
    \node[criminal, minimum size=0.95cm] at (11.1,2.85) {};
    \node[font=\scriptsize\bfseries, align=center, text=black!75] at (11.4,1.9) {Passive\\Observer};
    \foreach \y in {1.95,0.0,-0.75} \draw[densely dotted, red!55!black] (\sw,\y) -- (11.1,2.25);
    \draw[densely dotted, red!55!black] (\ms,1.30) to[out=25,in=195] (11.1,2.25);
    \node[idbox, fill=red!12, draw=red!55!black] (tei) at (11.2,0.85) {\textbf{TEI} --- 60-bit hardware serial};
    \node[idbox, fill=red!9,  draw=red!55!black] (issi) at (11.2,-0.25) {\texttt{ISSI}\,/\,\texttt{ITSI} --- permanent (\,$=$ \texttt{USSI} verbatim)};
    \node[idbox, fill=blue!12] (assi) at (11.2,-1.35) {\textbf{(V)ASSI} --- ephemeral alias, per LA};
    \node[idbox, fill=blue!7]  (evl) at (11.2,-2.35) {\texttt{Event Label} --- 10-bit, per call};
    \draw[<->, thick, black!60] (tei) -- node[right=1pt,font=\tiny,align=left]{\texttt{U-TEI}\\\texttt{PROVIDE}} (issi);
    \draw[<->, thick, black!60] (issi) -- node[right=1pt,font=\tiny,align=left]{migration\\reg. (1)(2)} (assi);
    \draw[<->, thick, black!60] (assi) -- node[right=1pt,font=\tiny,align=left]{\texttt{MAC-}\\\texttt{RESOURCE}} (evl);
  \end{tikzpicture}
  \caption{\textbf{TETRA registration links permanent, ephemeral, hardware, and event identities.}
    A migration exchange binds the home \texttt{ISSI}, exposed as the \texttt{USSI}, to the ephemeral \texttt{(V)ASSI}. \texttt{U-TEI PROVIDE} adds the hardware \texttt{TEI}, and \texttt{MAC-RESOURCE} links the \texttt{SSI} to a 10-bit event label. The \texttt{USSI} and event-label leaks persist in every security class. The \texttt{TEI} and general addresses are clear in class~1 and pseudonymized in classes~2 and~3~\cite{etsi-tetra}.}
  \label{fig:tetra-reg}
\end{figure*}

A TETRA subscriber is normally addressed over the air by a short alias---an
\texttt{ASSI}\,/\,\texttt{(V)ASSI}, whose pairing with the permanent
\texttt{ISSI}\,/\,\texttt{ITSI} the standard states ``cannot be derived from
knowledge of the \texttt{ITSI}'' and is ``only known to the network
operator''~\cite{etsi-tetra} (EN 300 392-1 \S7.2). Yet the binding of
\cref{fig:tetrapol-reg} is not TETRAPOL-specific: a passive observer recovers the
TETRA pairing directly from the registration exchange
(\cref{fig:tetra-reg}). In the migrating case (EN 300 392-2 \S16.4.1.1) the uplink
\texttt{U-LOCATION UPDATE DEMAND} is layer-2 addressed by the \texttt{USSI}, which
the standard requires the MS to form by \emph{copying its home \texttt{ISSI}
verbatim} (\S7.2.3, \S7.7.5). The downlink \texttt{D-LOCATION UPDATE PROCEEDING}
then returns the freshly assigned \texttt{(V)ASSI} and \texttt{MNI} as
\emph{mandatory} fields (Table 16.14). One round trip therefore binds the
permanent \texttt{ISSI} to the temporary \texttt{(V)ASSI} by PDU parsing
alone---, and, because the \texttt{USSI} is explicitly exempt from
the Encrypted Short Identity (ESI) mechanism (EN 300 392-7 \S4.2.6), even on an
encrypted class-2/3 cell.

Two further bindings complete the chain. If the base station sets the TEI-request
flag, the MS answers \texttt{U-TEI PROVIDE} carrying its 60-bit \texttt{TEI}---the
equipment serial---alongside its \texttt{SSI} in a single PDU (EN 300 392-7
Table A.28). The standard names this the ``\texttt{ITSI}\,/\,\texttt{TEI} mapping
at registration'' (\S5.4.1) and notes the \texttt{TEI} carries no cryptographic
seal. And every call grant that assigns a 10-bit \emph{event label} carries it
together with the \texttt{SSI} in one \texttt{MAC-RESOURCE} address element
(EN 300 392-2 Table 21.55, address type \texttt{101}$_2$). The event label is
likewise never encrypted (\S4.2.6), so once the observer resolves the
\texttt{SSI}, subsequent event-label-only bursts remain linkable.

The comparison to TETRAPOL is instructive. TETRA's \emph{steady-state}
pseudonymity is in fact \emph{stronger}: outside registration, ESI aliases the
\texttt{SSI} per location area under a rotatable common cipher key, a defence
TETRAPOL has no equivalent for. But the exposure concentrates precisely where the
protection is weakest---at registration and migration, where the \texttt{USSI}
carries the raw \texttt{ISSI}, and in the ESI carve-outs (\texttt{USSI}, event
label). On a security-class-1 cell, where no air-interface encryption applies
(EN 300 392-7 \S6.5.1), every one of these bindings travels in the clear, leaving
TETRA no better protected than TETRAPOL. The standard itself acknowledges the
risk---it \emph{recommends} (but does not require) that an MS ``not reveal its
true identity by sending \texttt{ISSI} in layer 3'' when an alias is available
(EN 300 392-2 \S16.4.1.1, NOTE 3), and that the \texttt{TEI} not be sent on
class-1 cells---so real-world exposure is deployment-dependent.

\subsection{Cryptographic Weaknesses: TEA1 and the Air-Interface Keystream}\label{sec:tetra-crypto}
TEA1 reduces its nominal 80-bit key to an effective 32 bits, making practical key
recovery possible. Related air-interface attacks can also recover keystream and
defeat protected identity signaling~\cite{meijer2023all}. These weaknesses turn
the metadata-derived targets of \cref{sec:tetra-meta} into potential content and
identity recovery, crossing the boundary that payload encryption is expected to
enforce. This result is specific to vulnerable cipher and protocol configurations and should not be generalized to every TETRA encryption algorithm.

\subsection{Location Privacy and Availability}\label{sec:tetra-loc}

In \emph{Analyzing TETRA Location Privacy and Network Availability}, Pfeiffer et al.~\cite{pfeifferAnalyzingTETRALocation2016} involved radio direction finding (RDF), but said effort
still relied on a signaling control plane design choice: the standard requires
coordination (``TETRA requires this regular signaling since MSes need to register at BSes,
authenticate, and exchange management data''\cite{pfeifferAnalyzingTETRALocation2016}) between a MS and its BS in configurable intervals, even when the MS is idle,
typically configured to be a short interval under 2 minutes.

This manifests in the
PHY layer as observable modulated signals enabling the aforementioned RDF. However, this
can be utilized for mere detection of near-field presence, through decoding of the exact
frames---and their sequencing in close temporal co-occurrence---employed for this purpose.
An observer both in the \emph{physical domain} and the \emph{RF domain} can leverage
such near-field fine granularity decoding to correlate, for example, a patrol vehicle
or officer to a specific on-air identity, even if temporary, and then extend the
inference to groups where the terminal appears active.

% ==================================================================
\section{P25 Security Analysis}\label{sec:p25}
% ------------------------------------------------------------------
\subsection{Unencrypted Control Channel and Identity Leakage}\label{sec:p25-meta}

P25 design exposes many operationally rich and important fields to adversarial inference.
This leakage is architectural: P25 voice encryption
protects the voice payload but explicitly does not protect radio identities on traffic
channels or call-request, assignment, registration, and affiliation information on a
trunked control channel. Data encryption has the same limitation for identities and
data-channel grants~\cite{tia102-aaab-a-1-2014}. Protecting those fields requires the
separate Air Interface Encryption (link-layer encryption) service, which the security
architecture defines as an optional overlay~\cite{tia102-aaab-a-2005}. As of PTIG's
July 2025 standards update, that service was still being drafted and the supporting
FDMA common-air-interface revision had only just been approved as a new
project~\cite{ptig-p25-standards-update-2025}.

A standard group-voice grant carries its
channel, group address, source address, emergency flag, protection flag, and priority. The protection flag applies to resources \emph{other than} the control channel, whose
default service option is ``not protected''~\cite{tia102-aabc-e-2019}. Registration and
affiliation messages similarly bind unit and group identities, while every Phase~1 data
unit exposes a 12-bit Network Access Code in its Network Identifier~\cite{tia102-baaa-a-2003}.
A passive observer can therefore recover who used which group, when, with what priority
and security state, and on which assigned channel even when the ensuing voice is
encrypted---the traffic-analysis risk identified by both TIA and Clark et
al.~\cite{tia102-aaab-a-2005,clark2011p25}.

\subsection{Weak / Optional Link Encryption and Clear Fallback}\label{sec:p25-crypto}

P25 security is an optional overlay rather than a uniform protocol property: systems may
mix clear and protected users and calls, and implementations may offer strong AES
alongside legacy DES and vendor-specific 40-bit RC4 modes~\cite{glass2012insecurity,tia102-aaab-a-1-2014,clark2011p25}.
The practical failure mode is therefore often operational rather than cryptanalytic.
In the radios studied by Clark et al., a commercial receiver configured for encryption still
accepted clear traffic, and the configuration-level remediation was not easy to find and deploy. A missing or desynchronized group key forced users back into
the clear, and ambiguous controls made that state easy to unwittingly select and hard for peers to
notice~\cite{clark2011p25}. Because the clear caller could still hear encrypted peers,
one radio could silently fail open while the rest of the group believed the exchange
remained protected. Their field study observed exactly this outcome in sensitive
federal-law-enforcement traffic. Strong payload cryptography cannot repair this
availability-driven clear fallback or the metadata exposure of \cref{sec:p25-meta}.
TETRAPOL presents a distinct clear-traffic risk because its ECH service is
normatively unencrypted (\cref{sec:tetrapol,sec:evaluation}).

\subsection{Affiliation Inference via Cipher Identification}\label{sec:p25-opsec-algid}

P25 implementations may offer privacy modes ranging from proprietary 40-bit RC4 to
DES-OFB and AES-OFB, as well as classified NSA Type~1 algorithms such as
BATON~\cite{glass2012insecurity,clark2011p25,wiki-baton}. The clear encryption-synchronization metadata
identifies the selected algorithm and key slot through the Message Indicator, ALGID,
and KID fields~\cite{tia102-baaa-a-2003}.

\Cref{tab:p25-type1-algid} lists a selection of these classified Type~1 ALGIDs, as
enumerated by the open-source OP25 decoder~\cite{op25}. Reading the ALGID from voice
headers and encryption-sync words lets a passive observer distinguish Type~1 from
commercial cryptography without recovering content. Combined with TG and
deployment context, that \emph{narrows the likely security domain and affiliation,} although
the ALGID alone does not identify an operator.

\begin{table}[t]
  \centering
  \caption{A selection of P25 Type~1 (classified) encryption algorithms, their
    in-the-clear ALGID codes~\cite{op25}, and approximate introduction years where
    publicly known.}
  \label{tab:p25-type1-algid}
  \small
  \begin{tabular}{@{}lll@{}}
    \toprule
    \textbf{ALGID} & \textbf{Type~1 algorithm} & \textbf{Introduced} \\
    \midrule
    \texttt{0x00}               & ACCORDION 1.3         & --- \\
    \texttt{0x01}/\texttt{0x41} & BATON (Auto Even/Odd) & ${\ge}\,$1995~\cite{wiki-baton} \\
    \texttt{0x02}               & FIREFLY               & c.\ 1987~\cite{wiki-stu3} \\
    \texttt{0x03}               & MAYFLY                & --- \\
    \texttt{0x04}               & SAVILLE               & late 1960s~\cite{wiki-saville} \\
    \texttt{0x05}               & PADSTONE              & --- \\
    \bottomrule
  \end{tabular}
\end{table}

% ==================================================================
\section{Cross-Protocol Synthesis}\label{sec:synthesis}
% ------------------------------------------------------------------

Across the
standards, an unprotected or under-protected signaling plane enables passive
longitudinal inference.
In \cref{sec:inference} we described five classes of inference. Here
we compare the protocol fields that instantiate these classes. \Cref{tab:matrix} gives the
overview, while the tables below map the concrete signaling messages. 
Protocol-specific cryptographic, implementation, and
operator weaknesses determine whether an adversary can also recover content.

\subsection{Operational Security and Confidentiality}\label{sec:synthesis-opsec}

Applying techniques from \cref{sec:inf-opsec}, \cref{tab:cross-protocol-synth-opsec} shows a
common invariant: each standard exposes enough cryptographic state to distinguish
clear from protected traffic and to separate key domains. The longitudinal
TETRAPOL instance is measured in \cref{sec:tp-ospec-patterns}. The P25 cipher-ID
case appears in \cref{sec:p25-opsec-algid}. These inferences do not require key
recovery or a breach of traffic-plane confidentiality.

\begin{table*}[t]
  \centering
  \caption{Confidentiality metadata carried in the clear.}
  \label{tab:cross-protocol-synth-opsec}
  \footnotesize
  \begin{tabularx}{\textwidth}{@{}l L{0.9} L{1.1}@{}}
    \toprule
    \textbf{Protocol} & \textbf{Carrying PDUs / codops} & \textbf{Fields exposed \& inference} \\
    \midrule
    TETRA & SYSINFO security-class broadcast, call-grant encryption mode, KSG/SCK signaling & security class (1/2/3), air-interface encryption on/off per call, cipher-key class, and clear ISSIs in class~1~\cite{meijer2023all} \\
    \addlinespace[2pt]
    TETRAPOL & codops carrying \texttt{KEY\_\allowbreak TYPE} / \texttt{KEY\_\allowbreak INDEX}, retransmitted encrypted-data codops & key type and key index in the clear, daily key-epoch rotation, and invariant ciphertext across same-epoch retransmissions (no randomized padding) \\
    \addlinespace[2pt]
    P25 & LDU/voice-header and call-grant \texttt{ALGID} / \texttt{KID} / Message Indicator (MI) & algorithm (cipher) ID, key ID, and MI in the clear, per-call clear-versus-encrypted state, and misconfiguration or clear-call detection~\cite{clark2011p25} \\
    \bottomrule
  \end{tabularx}
\end{table*}

\subsection{Topology, Geography, and Presence}\label{sec:synthesis-topology}

\Cref{tab:inf-topology} compares the broadcasts that make network discovery
self-guiding, and \cref{tab:inf-presence} compares the transactions that expose
registration, affiliation, and mobility. Their composition instantiates
\cref{sec:inf-topology,sec:inf-presence}: adjacent control frequencies extend the
survey, while timestamped site and group events associate activity with the
resulting graph. The TETRAPOL topology demonstration appears in
\cref{sec:tp-neighbor}, and cross-standard tracking is evaluated in
\cref{sec:eval-tracking}.

\begin{table*}[t]
  \centering
  \caption{Registration, affiliation, and mobility transactions.}
  \label{tab:inf-presence}
  \footnotesize
  \begin{tabularx}{\textwidth}{@{}l L{0.9} L{1.1}@{}}
    \toprule
    \textbf{Protocol} & \textbf{Carrying PDUs / codops} & \textbf{Fields exposed \& inference} \\
    \midrule
    TETRA & \texttt{U-LOCATION UPDATE DEMAND}, \texttt{D-LOCATION UPDATE ACCEPT}, ITSI attach/detach, \texttt{D-ATTACH/\allowbreak DETACH GROUP IDENTITY} & attach/detach timestamps, Location-Area moves, group-affiliation churn, and idle-device beaconing that aids localization~\cite{pfeifferAnalyzingTETRALocation2016} \\
    \addlinespace[2pt]
    TETRAPOL & \texttt{U\_REGISTRATION\_REQ}, \texttt{D\_REGISTRATION\_ACK/\allowbreak NAK}, \texttt{D\_TTI\_ASSIGNMENT}, cell-reselection codops & per-subscriber registration timing, diurnal PDU volume, per-TG activity, cell transitions \\
    \addlinespace[2pt]
    P25 & \texttt{Unit Registration} (\texttt{U\_REG\_REQ/\allowbreak RSP}), \texttt{Group Affiliation} (\texttt{GRP\_AFF\_REQ/\allowbreak RSP}), \texttt{Location Registration} (TSBK) & attach times, site roaming, affiliation changes over time~\cite{clark2011p25} \\
    \bottomrule
  \end{tabularx}
\end{table*}

\begin{table*}[tbp]
  \centering
  \caption{Signaling messages that expose network topology and geography.}
  \label{tab:inf-topology}
  \footnotesize
  \begin{tabularx}{\textwidth}{@{}l L{0.9} L{1.1}@{}}
    \toprule
    \textbf{Protocol} & \textbf{Carrying PDUs / codops} & \textbf{Fields exposed \& inference} \\
    \midrule
    TETRA & \texttt{D-NWRK-\allowbreak BROADCAST} (neighbor list), SYSINFO on BNCH (cell params), \texttt{SYNC} on BSCH & MCC/MNC, Location Area, serving/neighbor cell IDs and carrier frequencies, adjacency graph, and coverage mapping~\cite{meijer2023all} \\
    \addlinespace[2pt]
    TETRAPOL & \texttt{D\_SYSTEM\_INFO}, \texttt{D\_NEIGHBOURING\_\allowbreak CELL}, \texttt{D\_DDCH\_\allowbreak DESCRIPTION} & country / system\_id / network / product version, serving BS/RSW cell ID, control + DDCH frequencies, up to 12 neighbor cell IDs and their control-channel frequencies \\
    \addlinespace[2pt]
    P25 & \texttt{RFSS Status Broadcast}, \texttt{Network Status Broadcast}, \texttt{Adjacent Site Status Broadcast}, \texttt{Identifier Update} (TSBK) & WACN, System / RFSS / Site ID, adjacent-site control-channel frequencies, band plan~\cite{clark2011p25} \\
    \bottomrule
  \end{tabularx}
\end{table*}

\subsection{Organizational-Structure Inference}\label{sec:synthesis-org}

\Cref{tab:inf-org} maps the addressing, call-setup, paging, priority, and
emergency fields that instantiate \cref{sec:inf-org}. Pairing source/destination
identities with TG, priority, and emergency fields reconstructs the organizational
chart, chain of command, and unit function without decoding voice. Across the standards,
stable group addressing leaves membership, interaction, population, and activity
cadence observable even when subscriber identifiers rotate. Once a group is
associated with a real-world function, its presence alone can become operationally
sensitive. The P25 case study appears in \cref{sec:cs-p25}.

\begin{table*}[t]
  \centering
  \caption{Call-setup, paging, and priority signaling.}
  \label{tab:inf-org}
  \footnotesize
  \begin{tabularx}{\textwidth}{@{}l L{0.9} L{1.1}@{}}
    \toprule
    \textbf{Protocol} & \textbf{Carrying PDUs / codops} & \textbf{Fields exposed \& inference} \\
    \midrule
    TETRA & \texttt{U-SETUP}, \texttt{D-SETUP}, \texttt{D-TX GRANTED}, \texttt{D-CONNECT}, \texttt{D-ATTACH/\allowbreak DETACH GROUP IDENTITY}, \texttt{D/\allowbreak U-SDS-DATA} & source/destination SSI, TG membership, call priority, pre-emptive/emergency priority, status messages, and who-calls-whom \\
    \addlinespace[2pt]
    TETRAPOL & \texttt{D\_GROUP\_\allowbreak ACTIVATION}, \texttt{D\_GROUP\_PAGING}, \texttt{D\_CALL\_ALERT}, \texttt{D\_CALL\_CONNECT}, \texttt{D\_CONNECT\_DCH} & TG membership, \texttt{CALL\_PRIORITY}, \texttt{EMERGENCY\_TYPE}, who-pages-whom, and chain-of-command inference \\
    \addlinespace[2pt]
    P25 & \texttt{Group Voice Channel Grant}, \texttt{Unit-to-Unit Voice Channel Grant}, \texttt{Group Affiliation}, \texttt{Status/Message Update} (TSBK) & source/target unit ID, TG, call priority, emergency indicator~\cite{clark2011p25} \\
    \bottomrule
  \end{tabularx}
\end{table*}

% ==================================================================
\section{Evaluation}\label{sec:evaluation}
% ------------------------------------------------------------------
%
We evaluated the inference framework of \cref{sec:inference} against live,
operational deployments of all three standards, passively and receive-only, with no
key material. We had expected needing to stitch together our observations from many fragmentary exposures across long campaigns. Instead, \emph{a single decoded control carrier routinely bootstrapped an
entire deployment}, including its neighbor graph and frequency plan, its subscriber and
TG populations, their diurnal rhythm, and the key-rotation epochs of its
encrypted traffic. Where deployed, identity obfuscation merely degraded rather than defeated
this inference.

The difficulty lay in reach and reception: our limits were
capture geometry and the sampling constraints of remote receivers and the processing software (e.g., WH implements in software what is typically done by ASICs and dedicated processors for stateful and synchronous protocol operation), not the
protocols' defenses (when present at all). We used the following capture, sampling, and census methods to successfully overcome these constraints.

Radio enthusiasts commonly provide publicly available receivers connected to the Internet, as a community resource for radio exploration.  
We developed a framework (NOSYMESH) for leveraging these open receivers as a distributed system to validate our hypotheses about global deployments. For P25,
we implemented modifications to \texttt{dsd-neo}~\cite{dsdneo}, an open-source multi-protocol digital-voice decoder, for dumping P25 frames and their decoded fields,
and improved the polyphase resampler to tolerate low sampling rates (e.g., a P25 Phase 1 channel occupies
12.5\,kHz, requiring under 150\,ksample/s), a constraint imposed by the constellation of receivers
we used for observation (the default resampling required 1\,Msample/s). As a side-effect, our modification significantly
lowered the bandwidth requirements for parallel observation. We limited dwell time to ensure polite use of the open receivers, based on our understanding of the community practice.

We note that open receivers operated by hobbyists and ham radio operators could potentially be abused by a real adversary,
a passive observer that is not even located in the physical
vicinity of any of these systems.
Attackers' abuse of remote radios reportedly occurs in the wild: e.g., according
to~\cite{greenbergRussianSpiesJumped2024}, APT28's \emph{nearest neighbor} operation repurposed compromised
systems in adjacent buildings as unwitting Wi-Fi relays for the final physical hops to targets.
By analogy, an adversary could leverage any SDR-based platform it can access or modify against PMR systems---a class of remote-receiver abuse we anticipate
and term the \emph{nearest radio} attack.

\subsection{Metrics}\label{sec:eval-metrics}

We provide per-standard metrics for our data corpus in evaluation.
TETRA totals are GSMTAP packet records, TETRAPOL totals are decoder records, and
P25 totals are decoded event rows after exact-line deduplication. TETRA and TETRAPOL covered one public-safety deployment each, while P25 involved ten deployments.

\subsubsection{TETRA}

The TETRA corpus comes from one continuous downlink receiver, from one
public-safety deployment. It contains over 16 million packet records over 36.25~hours, of which $12,283,156$ decode as
TETRA PDUs, excluding malformed frames. The
mean record rate is 128.41~packets/s. The corpus exposes 251 distinct
clear nonreserved SSI operands, including 2,142
downlink location-update-family PDUs addressed to 230 SSI values.

\subsubsection{TETRAPOL}\label{sec:eval-metrics-tetrapol}

Our TETRAPOL corpus comprises 46 operational files with
over 33 million raw decoder records, from one deployment. They contain
12,154 TTI and 4,095 RTI address values, and 5,296,573 active-state emissions
covering 88 coverage-scoped group values. It also contains 1,586 ECH
activations, 250,558 system-information records, and 67,364 neighboring-cell
records. 877 non-encrypted ECH activations occurred.

\subsubsection{P25}

The P25 survey contains 688,866 records from ten systems. After deduplication of scheduler records, 600,036
unique events remain.

Three systems contribute 99.0\% of these events. A NIFOG interoperability deployment
contributes 95.2\% of all encrypted frames and 98.6\% of clear-RID frames.

Counting identities within their system namespaces yields 181
RIDs and 84 TGs. The bare numerical RID union is 178 because three values
collide across systems (subscriber IDs are often not unique across separate systems). Of 21,514 identity-bearing frames, 6,445, or 30.0\%,
expose the RID in clear. The NTIRN deep dive comprises 121,302 unique events over a sampled 58.5-hour span, with 177 RIDs, 75 TGs, and 610
crypto-classifiable voice calls.

\subsection{Quantitative Analysis of P25 Users}\label{sec:cs-p25}

\begin{figure}[th]
  \centering
  \includegraphics[width=\linewidth]{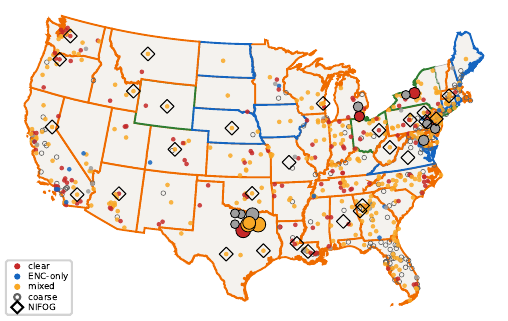}
  \label{fig:p25-conus}
\caption{\textbf{P25 census across CONUS.}
  Markers identify NIFOG-associated entries~\cite{cisa-nifog-2022} and observed systems. Large observed-system markers are scaled by decoded-frame count and colored for mixed (orange), clear-only (red), or encrypted-only (blue) traffic. State boundaries are also color coded with the same scheme. Locations are public database-sourced~\cite{radioreference}. Mixed deployments are prevalent.}
\end{figure}

Across our Continental US (CONUS) P25 survey, mixed clear-and-encrypted deployments are the
norm, and wherever subscriber and TG identities are transmitted they are carried in the
clear---from public-safety and federal users to utility and SCADA operators.

The National Interoperability Field Operations Guide (NIFOG)~\cite{cisa-nifog-2022},
published by the U.S.\ Cybersecurity and Infrastructure Security Agency (CISA), is a
public reference of interoperability channels and frequency assignments for
public-safety and government radio users. Deployment specifics---systems, sites, and
TG assignments---are further catalogued in public, enthusiast-maintained scanner databases, notably RadioReference~\cite{radioreference}.

Our census of 567 P25 systems across 48 states makes this concrete: only 19\% of their
roughly 138{,}000 talk groups are encrypted, just one state has at least half its talk groups
encrypted (median 18\%), and 27\% of systems carry no encryption at all.

In our survey, we
used these sources to look up
observed TG identities that were transmitted in the clear. These talk groups included law enforcement and public safety users, as well as SCADA operators. Our CONUS coverage
included the states of Texas, Ohio, Maryland, Virginia, and Washington, DC.

Our observations likely undercount existing traffic, 
due to the constraints
such as limited dwell times, connection reliability issues,
and proximity or capture geometry of the remote receivers.
Although our census aimed for a representative sample 
of exposed users, legacy ciphers, non-obfuscated
identities at scale, the actual exposure might be even larger than the collected data
demonstrates.

For Phase~2 (TDMA) deployments,
the only limiter was the ability to locate a reliable control channel carrier to obtain the parameters (e.g., WACN, SYSID, NACs) that, unlike Phase~1 (FDMA/C4FM), are not transmitted in the data or voice channels. Once these parameters are observed, the operator or organization can be identified, as well as the general class of operator (e.g., federal), with these identifiers being public in FCC and radio enthusiast databases.

\subsubsection{Law enforcement and public safety users}

In our observed census, one public safety deployment covering police and EMS users had its distinctive TG
operational details publicly available (via RadioReference~\cite{radioreference}). This included federal liaison, narcotics, special operations, counter-terrorism, and executive-protection divisions, each identified by its TG ID. The names were likely derived from the P25 alias or alpha tag transmitted in the clear.

A representative slice appears in \cref{tab:ncr-talkgroups}: each such function is carried as a plaintext TG identity regardless of whether its voice payload is encrypted.

\begin{table}[h]
  \centering\small
  \setlength{\tabcolsep}{5pt}
  \caption{Representative TGs on one National Capital Region public-safety system \cite{rr-sid7508}.}
  \label{tab:ncr-talkgroups}
  \begin{tabularx}{\columnwidth}{@{}r l X@{}}
    \toprule
    Dec & Alias tag & Name \\
    \midrule
    11139 & MPD EPU      & Executive Protection Unit \\
    11302 & MPD HS/CT    & Homeland Security / Counter-Terrorism \\
    11131 & MPD MND1     & Major Narcotics Division 1 \\
    11560 & MPD NSID 1   & Narcotics \& Special Inv. Div.\ 1 \\
    11041 & MPD SOD1 Evt & Special Operations Division 1 \\
    11117 & MPD OIA      & Office of Internal Affairs \\
    1039  & RIZ-16 CW-1  & MPD Citywide 1 \\
    145   & 0H4 MEDSTAR  & MedStar hospital \emph{(clear)} \\
    154   & 0H13 WHC     & Washington Hospital Center \emph{(clear)} \\
    728   & 011 EMS 1    & EMS 1, East \emph{(clear)} \\
    \bottomrule
  \end{tabularx}
\end{table}

Several of these groups, including most hospital and health-related services, were not visibly using encryption.

\subsection{Subscriber Tracking Across Standards}\label{sec:eval-tracking}

Across standards, we found that with sufficient observer density (e.g., more than one
receiver distributed across an area of interest), the failure to conceal subscriber and
system identity is a usable vector for tracking mobility across geographical zones. Once a
given TG's purpose is established, it becomes possible at scale to detect when it is operating
in different locations---especially sensitive for law enforcement users, who may depend on
the element of surprise and concealment. Under the perception that the system guarantees
confidentiality~\cite{NIJ2007VoiceEncryption,P25SteeringCommitteeCISA2023UserNeeds,Bundeswehr2021Tetrapol},
its users may inadvertently reveal both their affiliation and presence
without any breach of protected or encrypted content.

\subsection{TETRAPOL Emergency Call Extraction}\label{sec:tp-ech-extraction}

\begin{figure}[]
  \centering
  \includegraphics[width=\linewidth]{
    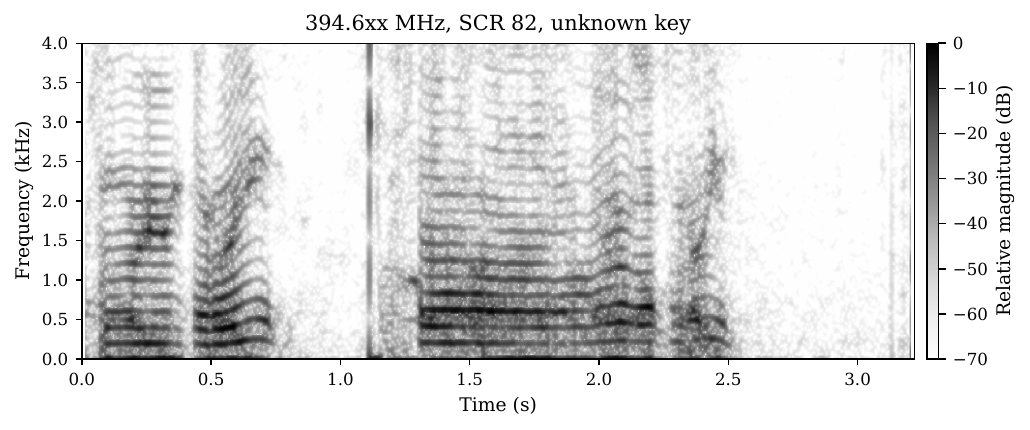
  }
  \caption{Spectrogram of a clear TETRAPOL emergency call (post-FEC and post-descrambling, and after RP-CELP decoding).}
  \label{fig:tetrapol-clear-call}
\end{figure}

The moment of greatest operational sensitivity is precisely the moment at which the protocol
removes content confidentiality: TETRAPOL specifies its Emergency Open Channel (ECH) service as
an unencrypted call.
PAS~0001-3-2~\cite{tetrapolPAS-0001-3-2} §4.4.33 requires the \texttt{D\_ECH\_ACTIVATION}
message to carry \texttt{KEY\_REFERENCE}~=~\texttt{UNENCRYPTED\_CALL} together with its
mandatory uplink and downlink dynamic-channel-scrambling parameters. The absence of traffic
encryption is thus written into the normative message definition, not merely an operator
configuration. (The clear-call mandate is not unique to ECH: §4.4.66 likewise applies to \texttt{D\_PRIORITY\_GRP\_ACTIVATION}.)
We confirmed this end to end: across five ECH activations in an operational deployment, all
five carried the clear-call key reference and yielded intelligible RP-CELP speech after
ordinary channel descrambling and error correction---without possession or recovery of any
cryptographic key---subject only to the publicly signaled dynamic-channel scrambling.
Figure~\ref{fig:tetrapol-clear-call} shows the spectrogram of one such call.

We obtained these calls by tracking the control plane. \texttt{D\_ECH\_ACTIVATION} is the
downlink message announcing an active emergency open channel. We implemented the explicit
grant-based tracking of \cref{sec:method-pipeline} in WH so that an observed activation---with
its advertised scrambling parameter (here \texttt{D\_CH\_SCRAMBLING}=82) and
\texttt{UNENCRYPTED\_CALL} key reference (\cref{sec:tp-ospec-patterns,app:tp-clear-classifier})---cues
retrospective extraction of the assigned traffic channel from IQ buffered before the grant was
seen, at whole-cell scale and without tuning to it in advance.

The recovered dialogue supplies semantic ground truth for the physical trigger. In one
representative call, dispatch informed the field operator that the terminal's emergency button
had been pressed, and the operator replied that it had been an accident---connecting a
distress-button press to the ECH activation, the assigned traffic channel, and the recovered
clear speech. Some of the five calls also showed a central dispatch checking on an operator's
well-being, and idiomatic cues identified the law-enforcement organization involved.

Because the observed grant mechanism depends on capture constraints (including processing time and
actual RF reception), we can only quantify exposure through the observed ECH activation records.

\subsection{Operational Security Observation via Metadata}

Besides segmenting traffic by key epoch and recognizing migration windows (Figure~\ref{fig:tp-keyday}),
we distinguished exceptional key domains, participating also in their own key rotation schedules, distinct from the network norm. This key domain separation provides inference of clearly confidentiality-level segregated groups of interest belonging to separate agencies or special units therein, whose data and voice traffic needs compartmentalization.

Additionally, because the keying and algorithm choices are signaled in the clear, we can observe which TGs permit mixed encryption and clear traffic
and which use salient cipher modes (\cref{sec:p25-opsec-algid,sec:tp-ospec-patterns}).

\subsection{Portable TETRAPOL Observation}\label{sec:cs-portabletetrapol}

We developed modifications for the MMDVM hotspot device to support TETRAPOL demodulation and decoding, and verified the implementation with IQ resampled from a live deployment in a lab-controlled environment with a HackRF SDR and direct interconnects.

\begin{figure}[]
  \centering
  \includegraphics[width=0.5\linewidth]{
    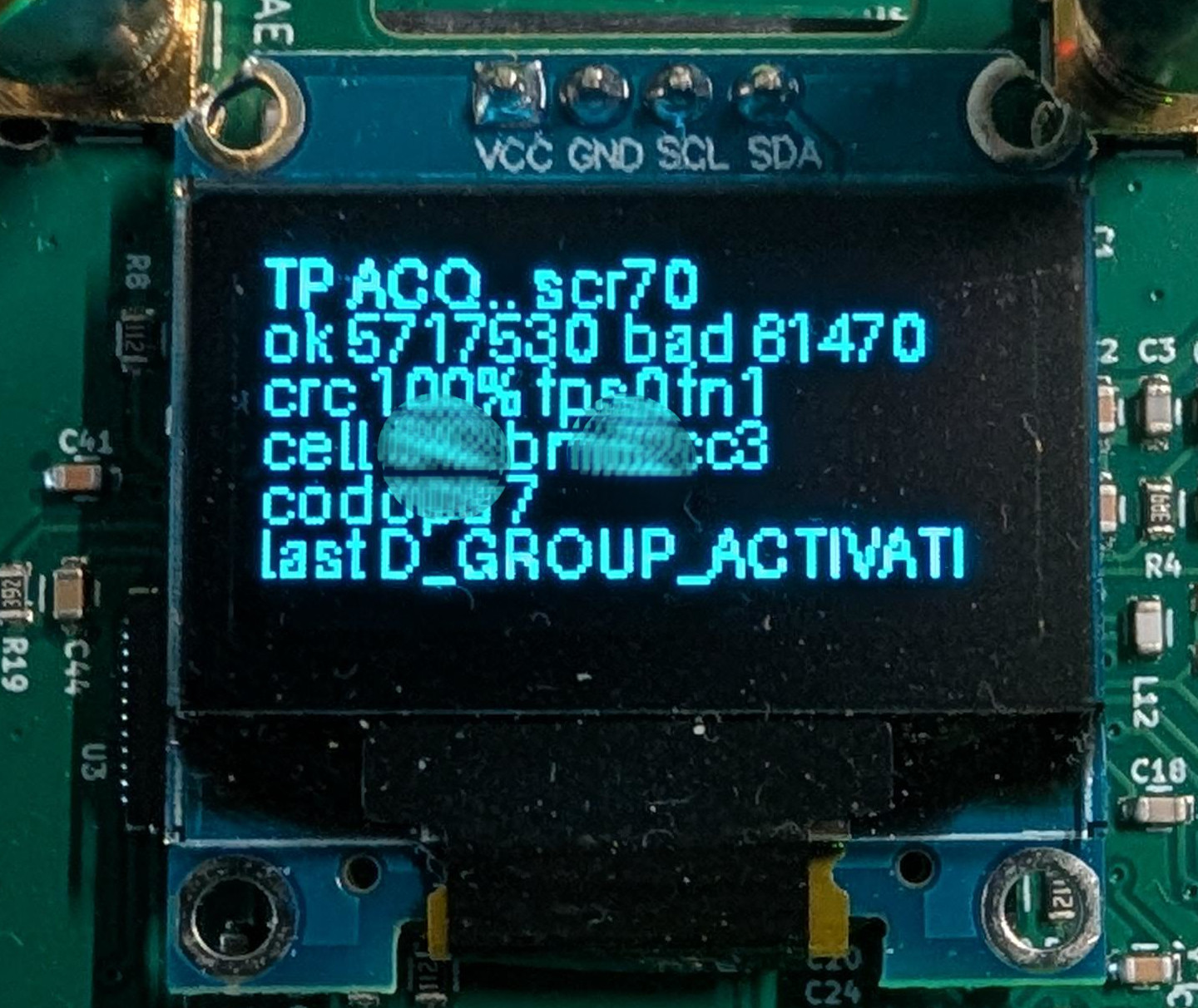
  }
  \caption{MMDVM-based TETRAPOL demodulation and decoding portable prototype showing active decoding of a \texttt{D\_GROUP\_ACTIVATION} downlink codop.}
  \label{fig:tetrapol-mmdvm}
\end{figure}

Despite the suboptimal TCXO of the low-cost hardware platform we utilized, our adjustments to the demodulation stage in firmware proved capable of sustaining real TETRAPOL decoding on a commodity MMDVM hotspot (Figure~\ref{fig:tetrapol-mmdvm})---portable, inexpensive interception is practical.

The platform is not a natural fit for TETRAPOL. Its downlink is GMSK ($h{=}0.5$, $BT{=}0.3$, ${\sim}8000$\,bit/s in a 12.5\,kHz channel), whereas the ADF7021's Gaussian shaper is fixed at $BT{\approx}0.5$ and its 14.7456\,MHz TCXO clocks only 7680\,bit/s---a ${\sim}4\%$ symbol-rate offset. We mitigated this by reusing the ADF7021's D-Star GMSK path, retiming its clock-recovery loop to the ${\sim}8000$\,bit/s rate and widening (clamping) the GMSK discriminator, enabling the receive-only chain to sustain live decoding.

% ==================================================================
\section{Related Work}\label{sec:related-work}
% ------------------------------------------------------------------

Glass et al.'s P25 research began in 2009 with OP25, the open-source P25 receiver,
running on a USRP software-defined radio, and therefore predates the better-known
USENIX study~\cite{glass2009sdr}. Their subsequent SecureComm work built on this
GNU Radio and USRP implementation to inspect and manipulate the physical and
data-link layers of P25~\cite{glass2012insecurity}. Most notably,
they reverse-engineered Motorola's Advanced Digital Privacy (ADP), confirmed its use
of RC4 with a 40-bit secret key, and showed how known silent vocoder codewords enable
practical exhaustive key recovery. The paper presents what appears to be the first
public description and practical cryptanalysis of ADP. This work is a direct
technical foundation for SDR-driven P25 analysis and deserves recognition as such.

Clark et al.~\cite{clark2011p25} provided complementary field evidence by surveying
US deployments with commercial radios. They documented clear law-enforcement calls
and selected metadata exposures, including the clear Network Identifier and an
apparent implementation error that exposed the Unit Link ID in encrypted voice
frames. We extend these foundations by composing signaling fields into identity,
topology, presence, organizational, and cryptographic-state inference across
protocols (\cref{sec:p25,sec:synthesis,sec:cs-p25}).

Meijer et al.~\cite{meijer2023all} analyzed TETRA's cryptographic primitives and
the consequences of designs that violate Kerckhoffs's Principle. Pfeiffer
et al.~\cite{pfeifferAnalyzingTETRALocation2016} examined location leakage and
network availability arising from the coordinated signaling between base stations
and subscriber radios. These works establish the cryptographic and traditional radio direction finding (location) risks,
but do not examine how clear control-plane fields compose over time to recover
network topology, presence, affiliation, and organizational structure. Our analysis
extends these results through signaling-field reconstruction and cross-protocol
inference (\cref{sec:tetra,sec:inference,sec:synthesis}).

For TETRAPOL, members of the \emph{brmlab hackerspace} in the Czech Republic
pioneered the first public prototype decoder for the standard. Its scope was mainly
control-channel decoding, and it did not implement many air-interface features needed
for whole-cell reconstruction. Jan Hrach subsequently developed the FCL channelizer
for the frequency-division monitoring requirements of standards such as TETRA and
TETRAPOL~\cite{hrachFrequencySpectrumMonitoring2016}. These efforts provide an
essential public tooling foundation. We extend it with broader protocol dissection,
longitudinal state reconstruction, and security inference
(\cref{sec:tetrapol,sec:evaluation}).

Dansarie et al.~\cite{10.1108/ICS-12-2024-0318} studied TETRA information security
from the users' perspective. Their work is valuable because LMR standards often lack
clearly stated security requirements, leaving trust expectations and operational
priorities to be inferred from practice. The authors deliberately exclude technical
vulnerabilities in protocols and implementations. Our work addresses that
complementary technical question and connects the resulting exposures to operational
consequences.

% ==================================================================
\section{Mitigations and Countermeasures}\label{sec:mitigations}
% ------------------------------------------------------------------

Our analysis demonstrates that, under the models where confidentiality of
communication is expected and relied on, there is no justification
for leaving group identifiers openly visible to passive observers.
In particular, subscriber, group, and topology identifiers must be protected from passive
observation. The adversary need not be particularly sophisticated or physically
proximal, as our P25 \emph{nearest radio} evaluation demonstrates (cf. also
\cite{greenbergRussianSpiesJumped2024}.) 

A layered key architecture can separate infrastructure from subscriber trust and
restrict topology broadcasts to the authorized nodes that require them. Subscriber
identifier rotation alone is brittle. Statistical linkage, long observation windows,
and physical correlation can reconnect an ephemeral identifier to a particular radio
or vehicle. Rotation should therefore be frequent and cryptographically unlinkable.
Padding and unique per-frame initialization vectors, including across retransmissions,
can also reduce known-length and ciphertext-correlation leakage.

The greater obstacle is the long replacement cycle
of public-safety communications systems,
whose deployments and contracts may span decades. Standards-level and implementation-level
mitigations will therefore propagate slowly.

% ==================================================================
\section{Discussion and Limitations}\label{sec:discussion}
% ------------------------------------------------------------------

Ethical and legal constraints most sharply limited this work. Despite strong
indications that active attacks and deeper cryptanalysis merit study, we did not pursue
active attacks or attempt a cryptographic break using live TETRAPOL deployment data.
We also lacked a controlled base station. % or network laboratory.
Continuous legal and
ethical review of passive observations substantially narrowed the work.

Capture geometry imposed a second limit. TETRA and TETRAPOL deployments may carry
user-defined applications, including SDS. Our observations of exposed application data
indicate a sensitive attack surface, but insufficient access and proximity kept a systematic
study of that surface out of reach.

We hope
this work draws further academic attention and fosters open collaboration among
stakeholders, vendors, and interested parties.

% ==================================================================
\section{Conclusions and Future Work}\label{sec:conclusion}
% ------------------------------------------------------------------

Across TETRA, TETRAPOL, and P25, the structural split between an unprotected
signaling plane and (often optional) encrypted traffic plane makes passive, protocol-aware metadata
inference a general attack. Standard-specific weaknesses intensify it and can cross the
threshold from \emph{knowledge of communication} to \emph{inference of content}.
\texttt{WH} composes grants, channel assignments, activity, and frame-level semantics
to cue downlink and uplink demodulation, decoding, and tracking. The exposed signaling
reveals when, where, and what to observe while reducing the required computation and
hardware.

Even without content recovery, this leakage reveals organizational structure,
presence, mobility, shift turnover, security posture, and preparedness. The threat is
particularly acute for law enforcement and first responders. Obscurity offers, at
most, a temporary barrier. Affordable SDRs and public tooling have
removed the historical cost and expertise assumptions on which it depended.

Traffic analysis remains difficult to prevent in radio protocols. Our future work
includes firmware analysis, controlled active-attack experiments, rehosting hardware
\emph{black-box} cryptographic modules, developing open-source drop-in cryptographic
coprocessors, and continued cooperation with stakeholders and users.

% ==================================================================
% ------------------------------------------------------------------
\bibliographystyle{IEEEtran}
\bibliography{witchhunt-public-safety-comms-1}

% ==================================================================
% ------------------------------------------------------------------
\appendices
% ------------------------------------------------------------------
\

% ==================================================================
\section{Ethical Considerations and Responsible Disclosure}\label{app:ethics}
% ------------------------------------------------------------------
All measurements in this work were passive and receive-only. We transmitted
nothing, mounted no active (e.g., denial-of-service) attacks, and never authenticated
to or joined a network. Our analysis targets the signaling plane and elements
of the management and traffic planes, which these
systems already broadcast unencrypted. We recovered data traffic only to the minimum needed to
demonstrate the weakness. For the emergency-call analysis we processed voice content only to
establish intelligibility, encryption state, and the semantic relationship between the observed
ECH activation and the terminal's emergency control. We retained no transcript beyond the
minimal excerpt needed to establish that relationship, and publish no names or identifying
operational content.

We treat all captured identifiers as sensitive. Subscriber and unit IDs,
cell, network, and location-area identifiers, exact carrier frequencies, and recovered
coordinates are anonymized or coarsened in the public copies of the paper---carrier
frequencies masked to band level, identifiers consistently relabeled---while the true
values are withheld, and we release no raw captures, recovered keys, third-party audio, or
operator-identifying data. All reception was conducted in accordance with
applicable law.

We withhold any tooling that would enable trivial misuse or endanger users.
Any tooling under limited and justified dissemination is unrestricted fundamental research intended for publication and defensive evaluation, for
reproducing our results on the researcher's own equipment. It confers no
capability beyond that already available to a determined passive adversary.
We likewise omit implementation details whose disclosure would ease replication of the more sensitive capabilities without adding to understanding of the underlying weakness.
Preprints and final revisions are published with best-effort harm reduction and in the public interest.

\section{TETRAPOL Clear-versus-Encrypted Call Classification}
\label{app:tp-clear-classifier}

WH assigns each TETRAPOL call one of two clear/encrypted verdicts, and the
difference between them governs how the label may be used. The
\emph{signaling verdict} $\widehat S_{\mathrm{sig}}$ is \emph{authoritative}: it
is read directly from the standard-mandated \texttt{KEY\_REFERENCE} field of a
fully validated call grant, and it is the only verdict that labels a call clear
or encrypted. It is what drives the clear-call detection behind the emergency
call extraction of \cref{sec:tp-ech-extraction}, where an
\texttt{UNENCRYPTED\_CALL} key reference cues retrospective recovery of clear
voice. The \emph{advisory verdict} $\widehat S_{\mathrm{adv}}$ is a heuristic
fallback used only when no grant supplies a key reference: it scores the decoded
voice and can report a call \emph{candidate clear}, \emph{not clear}, or
\emph{inconclusive}, but it never overrides $\widehat S_{\mathrm{sig}}$ and never
asserts ``encrypted.'' Its sole purpose is to surface clear speech on a channel
captured without a decodable grant.

\subsection{Signaling verdict (authoritative)}

For a complete, structurally valid grant or channel-activation PDU, let $K$ be
the decoded \texttt{KEY\_REFERENCE} octet. Its high nibble is the key type
$T=K\mathbin{\gg}4$, and its low nibble is the key index
$I=K\mathbin{\&}\mathtt{0x0f}$. PAS~0001-3-2~\cite{tetrapolPAS-0001-3-2},
\S5.3.41, yields the implemented verdict
\begin{equation}
  \widehat S_{\mathrm{sig}}(K)=
  \begin{cases}
    \text{clear}, & I=0 \ \wedge\ T\in\{0,15\},\\
    \text{encrypted}, & \text{otherwise}.
  \end{cases}
  \label{eq:tp-signaling-classifier}
\end{equation}
Equivalently, only $K\in\{\mathtt{0x00},\mathtt{0xf0}\}$ is clear. Every
other decoded value is encrypted, including $\mathtt{0x50}$, whose zero index
does not make its OAK key type clear. An absent key reference remains unknown
and is never replaced with zero.

Before applying \cref{eq:tp-signaling-classifier}, the decoder validates the
complete PDU, including its mandatory tail, reserved fields, defined
enumerations, and exact message consumption. The frame check sequence is not a
discriminator, since ciphertext is equally FCS-valid. A byte sequence that
merely reaches the key-reference position is therefore not authoritative.

The verdict is then cached per \emph{allocation}, identified by the carrier
together with its locked dynamic-scrambling constant. A known verdict survives a
grant timeout (the grant time-to-live default is $30$\,s) only within the same
carrier-and-scrambling allocation: while the carrier holds the same scrambling,
an interval with no fresh grant reuses the cached verdict rather than reverting
to unknown. A scrambling change or release, a conflicting decoded verdict, a
lost lock, or a call-ending idle gap ($\approx\!6$\,s without voice) retires the
allocation and starts a new one with unknown state. Scoping the verdict to the
scrambling constant keeps a stale label from leaking across calls.

\subsection{No-grant speech classification (advisory)}

When no grant supplies $K$, WH scores the decoded voice for a speech-like
signature. Clear and encrypted TETRAPOL voice differ sharply in their
short-term statistics: a genuine speech envelope evolves smoothly from frame to
frame and its quantizer indices occupy a characteristic mid-entropy band,
whereas ciphertext yields near-independent coefficients with no frame-to-frame
correlation and near-uniform indices. The test reduces a call to three summary
features---an envelope \emph{coherence} $C$, a width-normalized index
\emph{entropy} $H$, and a \emph{distinct-block fraction} $D$---and accepts only
the clear-speech region of $(C,H,D)$. Treating mean, variance, covariance, and
Shannon entropy as standard estimators, the procedure is

\begin{lstlisting}[language=Python]
# advisory clear-speech verdict  S_adv
# in:
#   B[1..N] = CRC-3-valid 120-bit RP-CELP blocks
# const:
#   Nmin=8, eps=1e-12,
#   tauC=0.15, Hlo=0.80, Hhi=0.97, tauD=0.60
if N < Nmin:
    return INCONCLUSIVE

# each log-area-ratio (LAR) index
for i in 1..10:
    r_i  = [reflection(B[n], i) for n in 1..N]
    v    = var(r_i)
    # lag-1
    a[i] = 0 if v <= eps else cov(r_i[1:N-1], r_i[2:N]) / v

# envelope coherence
C = clip(mean(a[1..10]), 0, 1)

# width-normalized
H = mean(entropy(x[:, i]) / b[i] for i in 1..10)

# distinct-block fraction
D = num_distinct(B[1..N]) / N

if C >= tauC and Hlo <= H <= Hhi and D >= tauD:
    return CANDIDATE_CLEAR

return NOT_CLEAR
\end{lstlisting}

\noindent The formulas below fix the exact estimators the pseudocode treats as
library calls. Given $N$ CRC-3-valid, 120-bit RP-CELP codec blocks $B_n$, for
each of the ten log-area-ratio (LAR) indices the decoder reconstructs a
reflection coefficient $r_{n,i}$. It then computes the mean $\bar r_i$, variance
$v_i$, lag-one covariance $q_i$, autoregressive coefficient $a_i$, and mean
coherence $C$ as
\begin{align}
  \bar r_i &= \frac{1}{N}\sum_{n=1}^{N}r_{n,i},\\
  v_i &= \frac{1}{N}\sum_{n=1}^{N}(r_{n,i}-\bar r_i)^2,\\
  q_i &= \frac{1}{N-1}\sum_{n=1}^{N-1}(r_{n,i}-\bar r_i)(r_{n+1,i}-\bar r_i),\\
  a_i &=
  \begin{cases}
    0, & v_i\leq 10^{-12},\\
    q_i/v_i, & v_i>10^{-12},
  \end{cases}\\
  C &= \operatorname{clip}_{[0,1]}
       \left(\frac{1}{10}\sum_{i=1}^{10}a_i\right).
  \label{eq:tp-coherence}
\end{align}
The coherence $C$ is the mean lag-one autocorrelation of the ten
reflection-coefficient tracks, clipped to $[0,1]$; it measures how smoothly the
spectral envelope evolves, and is high for real speech but near zero for
ciphertext. The variance floor $10^{-12}$ guards the ratio $q_i/v_i$: a
coefficient whose reconstructed track is essentially constant carries no
autocorrelation evidence, so it contributes $a_i=0$ rather than a ratio
dominated by floating-point noise.

Let $x_{n,i}$ be LAR index $i$ of block $n$, with quantizer field widths
$\boldsymbol b=(5,5,4,4,4,3,3,3,3,3)$ bits, and let
$p_{i,j}=\frac{1}{N}\,\#\{\,n:x_{n,i}=j\,\}$. The width-normalized index entropy
$H$ and the distinct-block fraction $D$ are
\begin{align}
  H &=\frac{1}{10}\sum_{i=1}^{10}\frac{1}{b_i}
       \left(-\!\!\sum_{j:p_{i,j}>0}\!\! p_{i,j}\log_2 p_{i,j}\right),\\
  D &=\frac{|\{B_1,\ldots,B_N\}|}{N}.
  \label{eq:tp-advisory-features}
\end{align}
The inner sum is the base-2 Shannon entropy of index $i$'s empirical
distribution, since a $b_i$-bit field carries at most $b_i$ bits of entropy,
dividing by $b_i$ expresses it as an occupancy fraction in $[0,1]$ and renders
the ten indices---which have different alphabet sizes---comparable before
averaging. Real speech occupies a mid band ($H\approx0.89$--$0.93$), whereas
near-uniform ciphertext indices give $H\approx1$. The distinct-block fraction
$D$ is the share of the $N$ blocks $B_n$ that are unique: real speech is almost
entirely distinct, whereas a degenerate reach-back or idle alias---the decoder
latching a fixed-scrambling carrier that is idle or replaying a handful of frames from a prior call---repeats blocks and drives $D$ down.

With the chosen thresholds, define the acceptance gate
\begin{equation}
  \mathcal G=[C\geq0.15]\wedge[0.80\leq H\leq0.97]
             \wedge[D\geq0.60],
\end{equation}
so that the advisory verdict is
\begin{equation}
  \widehat S_{\mathrm{adv}}=
  \begin{cases}
    \text{inconclusive}, & N<8,\\
    \text{candidate clear}, & N\geq8\ \wedge\ \mathcal G,\\
    \text{not clear}, & N\geq8\ \wedge\ \neg\mathcal G.
  \end{cases}
  \label{eq:tp-advisory-classifier}
\end{equation}
Each bound in $\mathcal G$ separates observed clear speech from a specific
confound. Clear speech has smooth envelopes ($C\approx0.62$--$0.68$), mid-band
entropy ($H\approx0.89$--$0.93$), and mostly distinct blocks. Encrypted blocks
have $C\approx0$ and $H\approx1$. The coherence floor $0.15$ thus sits far below
clear speech yet well above ciphertext. The entropy band $[0.80,0.97]$ excludes
both near-uniform ciphertext ($H\approx1$) and the low-entropy degenerate
aliases ($H\approx0.26$), and $D\geq0.60$ rejects their repeated blocks
($D\approx0.20$). A reach-back can fake high coherence through repeated frames,
but its low entropy and low distinct fraction reject it. Requiring $N\geq8$
blocks keeps the estimates stable. Because the test only confirms or fails to
confirm \emph{clear} speech, \emph{not clear} withholds a clear label rather
than asserting encryption---the encrypted judgment is reserved for the
authoritative $\widehat S_{\mathrm{sig}}$.

\section{TETRAPOL Call Post-FEC Spectral Rendering}\label{app:tetrapol-calls-spectral}

\begin{figure}[h]
  \centering
  \includegraphics[width=\linewidth]{
    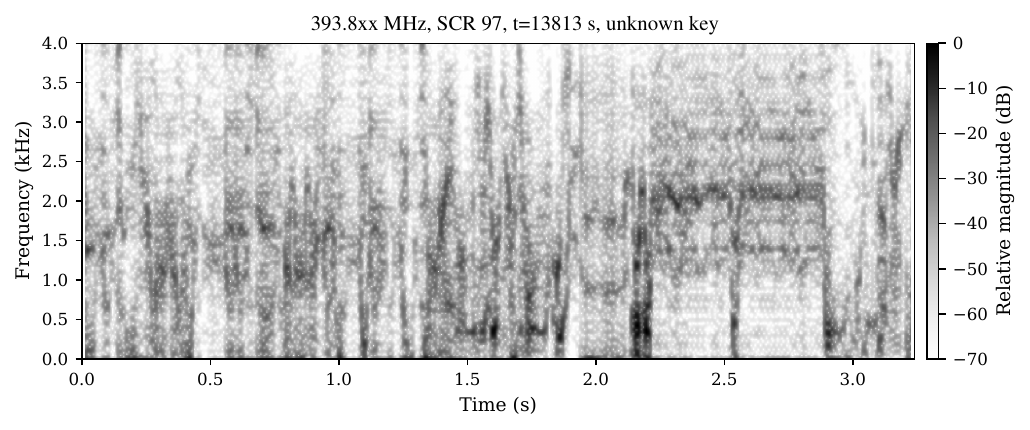
  }

  \includegraphics[width=\linewidth]{
    figs/tetrapol_clear_ech_spectrogram.pdf
  }
  \caption{Spectrograms of encrypted call and clear TETRAPOL emergency calls  (post-FEC and post-descrambling).}
  \label{fig:tetrapol-clear-vs-enc-calls}
\end{figure}

WH preserves no-grant streams with an unknown signaling state for later
analysis. The advisory RP-CELP signature in \cref{app:tp-clear-classifier}
operates on these streams without overriding the signaling verdict. Clear and
encrypted calls exhibit the spectral differences shown in
\cref{fig:tetrapol-clear-vs-enc-calls}.

\subsection{Spectral Rendering}

The WH pipeline classified the call from its ECH grant and retrospectively
extracted it from the ring buffer while preserving the original IQ:

\begin{lstlisting}
20881874:[ECH-DL #1] channel_id=2902 dl_carrier=3946xxxxx d_scr=82 u_scr=76 t=2164.7s codop=0x56 key_reference=0x00 [CLEAR-EMERGENCY]
20884669:[ECH-DL-CAPTURE] ch2902 c3946xxxxxHz codop=0x56 d_scr=82 scr_used=82 FIXED=true key_reference=0x00 clear=true t_grant=2164.7s extract=132000000 samples (55.0s) frames=2749 voice=177 voice_crc=167/177 wav=(...)/ech_dl_c3946xxxxxHz_ch2902_dscr82_crc167of177_t2165s.wav iq=(...)/ech_dl_c3946xxxxxHz_ch2902_t2165s.cs16
\end{lstlisting}

\section{Hardware}\label{app:sdr-hw}

\subsection{Self-built Experimental Multi-SDR Platform}\label{app:ghettosdr}

\begin{figure}[h]
  \centering
  \includegraphics[width=\columnwidth]{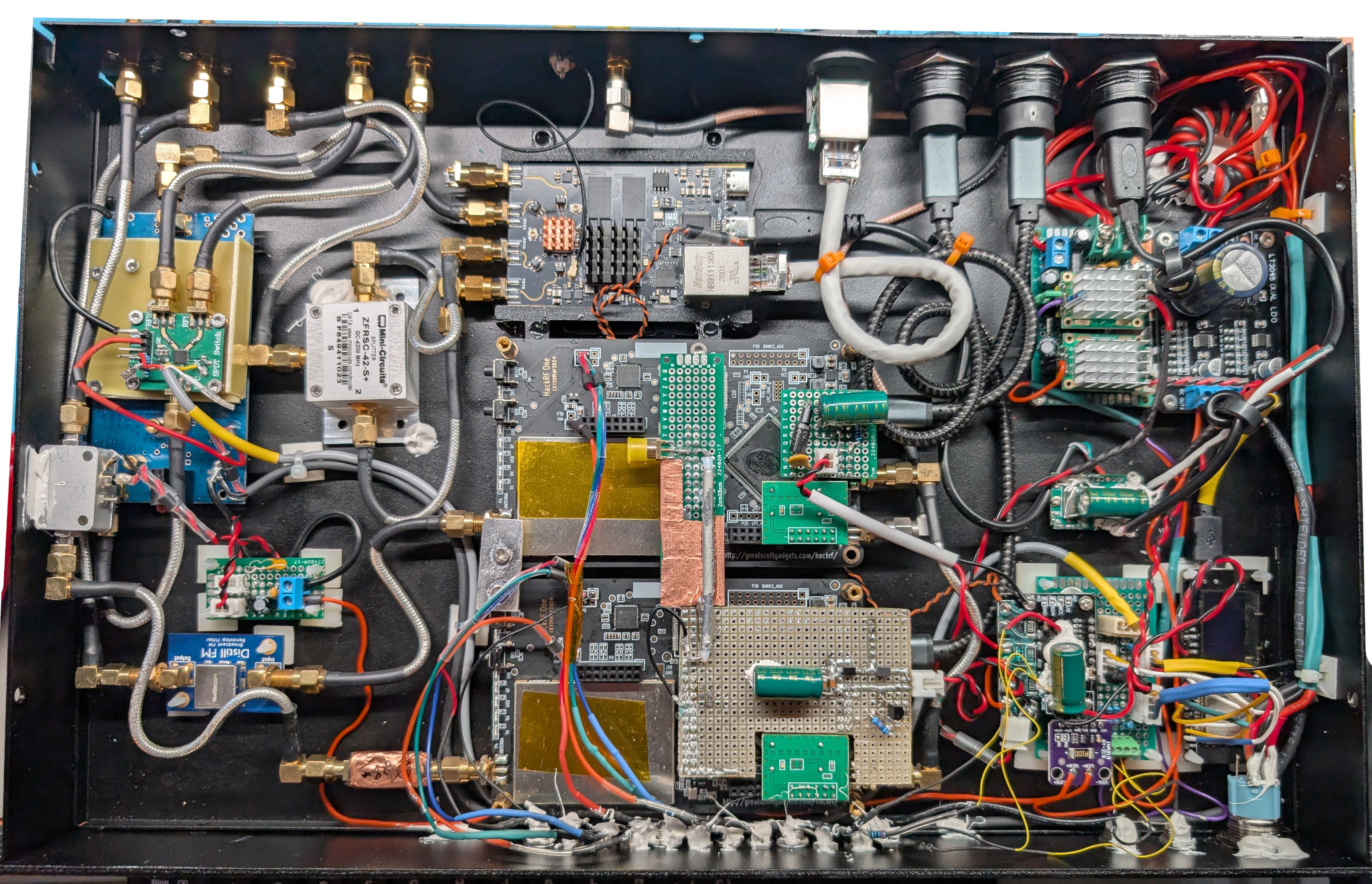}
  \caption{Self-built Experimental Multi-SDR Platform}
  \label{fig:ghettosdr}
\end{figure}

The SDR hardware employed throughout this effort is pictured in Figure~\ref{fig:ghettosdr}.
The design is not particularly sophisticated: power is supplied by a regulated linear supply,
converted to a stable ~5V DC by ultra-low noise LDO regulators (TPS7A4700 and LT3042), and monitored in a closed
loop by a controller that can manage individual supplies to each SDR. On the RF side,
an ad-hoc selection of components enables filtering of out-of-band strong signals
(primarily commercial FM broadcast), limiting of strong RF signals, power for antenna-feed
low noise amplifiers (decoupled from the SDRs) and antenna switching with a GaAs RF SPDT switch
(Analog Devices HMC349ALP4CE).

The two HackRF boards are modified (\cref{app:hackrf-vbus-mod}) and share a common TCXO.
The Pluto-design SDR has a clock input for an external GPS disciplined oscillator (not
pictured). A PPS pin is broken out for both HackRFs for optional coherent reception.

Reception used colinear and discone antennas behind bandstop filtering that also
rejected out-of-scope RF energy. A spare SDR transmit port doubles as a
built-in path for sensitivity checks and feedline/antenna diagnostics, and a
separate BladeRF/HackRF bench inside a shielded RF-isolation cage supported onsite
reverse engineering.

The system was designed, built and fielded in under two weeks time, and was mostly
built from a \emph{parts bin}, to demonstrate that the cost and skill barrier today is
orders of magnitude lower than it used to be.

\subsection{HackRF VBUS Isolation PCB Modification}\label{app:hackrf-vbus-mod}

The dual HackRF configuration with a shared TCXO necessitated preventing any
remnant---radiated or conducted---noise from impacting function. Despite the
use of ultra-low noise regulators (Texas Instruments TPS7A4701), this was a
desirable improvement. The VBUS trace was cut off under microscope, with the
ferrite bead removed. The trace was then cut in two spots, allowing placement
of a diode to prevent backfeed from the independent pin-fed 5V power towards
the USB host, and the original ferrite immediately after.

\begin{figure}[h]
  \centering
  \includegraphics[width=\columnwidth]{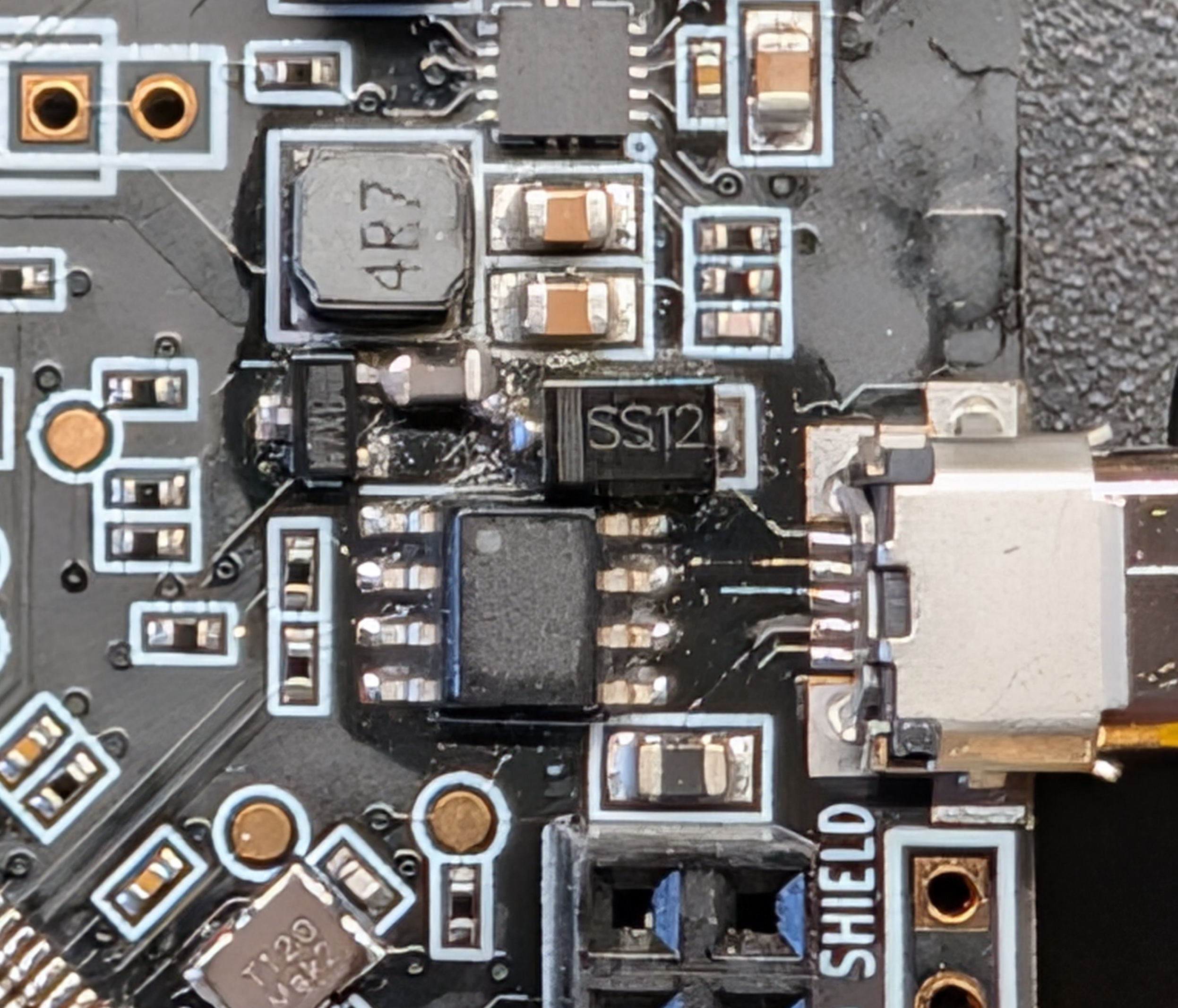}
  \caption{HackRF One with USB \texttt{VBUS} isolated for the dual-receiver platform in \cref{sec:method-pipeline}.}
  \label{fig:hackrf-vbus-mod}
\end{figure}

The modification requires a steady hand and microscope magnification.

\end{document}